\documentclass{aastex}
\newif\ifgrayscale
\grayscaletrue % to compile with grayscale images
\newif\ifincludeplots
\includeplotstrue %uncomment to compile with figures. you also need to comment out the line below.
\newcommand{\sameyear}[1]{}

\newcommand{\eg}{{\it e.g.}}

\newcommand{\etal}{{\it et~al.}}

\newcommand{\x}{ \; \mathrm{x} \;}

\newcommand{\mps}{\,\mathrm{\frac{m}{s}}}

\newcommand{\gpcmc}{\,\mathrm{\frac{g}{cm^3}}}

\newcommand{\au}{\,\mathrm{au}}
\newcommand{\km}{\,\mathrm{km}}

\newcommand{\myr}{\,\mathrm{Myr}}

\newcommand{\pv}{p_V}

\newcommand{\tdeg}{^{\circ}}

\newcommand{\invkm}{\mathrm{km^{-1}}}
\begin{document}

\title{Yarkovsky V-shape identification of asteroid families}

\author{
Bryce T. Bolin\altaffilmark{1} (bryce.bolin@oca.eu),
Marco Delbo\altaffilmark{1},
Alessandro Morbidelli\altaffilmark{1},
Kevin J. Walsh\altaffilmark{2}
}

\received{June 20, 2016}
\accepted{September 20, 2016}

\slugcomment{74 Pages, 17 Figures}
\altaffiltext{1}{Laboratoire Lagrange, Universit\'e C\^ote d'Azur, Observatoire de la C\^ote
d'Azur, CNRS, Blvd. de l'Observatoire, CS 34229, 06304 Nice cedex 4,
France}
\altaffiltext{2}{Southwest Research Institute, 1050 Walnut St. Suite 300, Boulder, CO 80302, United States}

\shorttitle{Identifying Asteroid Family Yarkovsky V-shapes}
\shortauthors{Bolin \etal}

% ABSTRACT -----------------------------------------------

\begin{abstract}
There are only a few known main belt (MB) asteroid families with ages greater than 2 Gyr \cite[][]{Broz2013b, Spoto2015}. Estimates based on the family producing collision rate suggest that the lack of $>2$ Gyr-old families may be due to a selection bias in current techniques used to identify families.  Family fragments disperse in their orbital elements, semi-major axis, $a$, eccentricity,  $e$, and inclination, $i$, due to secular resonances, close encounters with massive asteroids and the non-gravitational Yarkovsky force. This causes the family fragments to be indistinguishable from the background of the main belt making them more difficult to identify with the hierarchical clustering method (HCM) with increasing family age. The discovery of the Eulalia and new Polana  families in the inner belt relied on new techniques because Yarkovsky spreading made them too disperse to be identified using the classical HCM. The techniques used to discover the new Polana  and Eulalia families are modified here to identify asteroid families by searching for correlations between $a$ and asteroid diameter, $D$, or absolute magnitude, $H$. A group of asteroids is identified as a collisional family if its boundary in the $a$ vs. $\frac{1}{D}$ or $a$ vs. $H$ planes has a characteristic V-shape which is due to the size dependent Yarkovsky spreading. The V-shape boundary is identified with two separate techniques. The first technique identifies a border by measuring a steep drop between the number of objects inside and outside of the border. The second technique identifies the V-shape border by measuring a peak in the number density of objects in $a$ vs. $\frac{1}{D}$,$H$ space. Families are identified with just one or both V-shape identifying techniques. The V-shape techniques are demonstrated on the known families of Erigone, Vesta, Koronis, and families difficult to identify by HCM such as Flora, Baptistina, new Polana, Eulalia and Karin. Future applications of the technique, such as in a large scale search for $>\;2$ Gyr-old families throughout the MB, are discussed. 
\end{abstract}


\maketitle

% RUNNING HEAD & CORRESPONDENCE -----------------------------------------------

\clearpage

\mbox{ }

\vspace{0cm}

%\begin{center}
\noindent {\bf Proposed Running Head:} Yarkovsky V-shape identification of asteroid families \\
%\end{center}

\vspace{10cm}

\noindent {\bf Editorial correspondence to:} \\
Bryce Bolin \\
Observatoire de la Cote d'Azur \\
Boulevard de l'Observatoire \\
CS 34229 \\
06304 Nice, France \\
Phone: +33 04 92 00 30 81 \\
Fax: +33 (0) 4 92 00 30 33 \\
E-mail: bbolin@oca.eu

% INTRODUCTION -----------------------------------------------------------

\section{Introduction}
\label{s.Introduction}

Asteroid families are formed during collisional catastrophic disruption and cratering event excavations of larger parent bodies \citep[][]{Michel2001, Michel2003, Michel2015}. Although dispersed in space, the family members typically have proper orbital elements, semi-major axis $(a)$, eccentricity $(e)$ and inclination $(i)$, close to that of the parent body \citep[][]{Hirayama1918, Nesvorny2015a}.
 
It is generally believed that the collision rate among asteroids in the Main Belt remained relatively constant during the last $\sim$4 Gyr \citep[][]{Bottke2005a}. Thus, asteroid families should have been produced roughly uniformly over time, with a frequency dependent on the collisional lifetimes of the parent bodies, i.e. on their size \citep[][]{Bottke2005b, Broz2013b}. However, a systematic study of the ages of the known asteroid families shows a deficit of families with ages larger than 2 Gy for all parent body sizes \citep[][]{Nesvorny2005,Broz2013b,Spoto2015,Carruba2016c}, in contrast with the expectation of a roughly constant production rate. If this was true, it would imply an unexpected collisional history of the asteroid belt, with a steep increase in the mutual collision rate in the last few Gy.
 
Before reaching this strong conclusion, however, one has to address the bias against the identification of the oldest families. Asteroid families are usually identified as statistically significant clumps of bodies in the space of proper elements ($a$,$e$,$i$) \citep[]{Milani1994a, Knezevic2002, Knezevic2003, Nesvorny2015a}. The identification is typically done with the so-called Hierarchical Clustering Method \citep[HCM;][]{Zappala1990,Bendjoya2002,Milani2014}. HCM in its simplest form identifies families by measuring the relative velocity between asteroids' proper $a$,$e$,$i$ and a central reference asteroid and selecting all asteroids below a cutoff value in velocity. The cutoff value in relative velocity is determined by comparing the actual number of asteroids in the velocity cutoff to the number of asteroids in the cutoff that have quasi-random distributed elements. Families are considered statistically significant if their number exceeds the quasi-random level for a given cutoff velocity. Alternative methods have been tested \citep[e.g., the Wavelets method:][]{Bendjoya1991}, which give similar results in identifying asteroid families compared to HCM. Attempts to identify asteroid clusters in the space of proper orbital frequencies, $n$, asteroid mean-motion, $g$, secular frequency of pericenter and $s$, secular frequency of node, similar to classical HCM have also been performed \citep[][]{Carruba2007}. 

All these methods encounter the problem that asteroid families disperse over time. The proper semi-major axis changes for all asteroids due to the so-called Yarkovsky effect \citep[][]{Farinella1998, Bottke2001,Nesvorny2002a,Vokrouhlicky2015}. This is a non-gravitational effect due to the non-zero thermal inertia of the surface of the asteroid so that the emission of thermal radiation by a rotating asteroid illuminated by the Sun occurs preferentially in a direction offset relative to the Sun-asteroid line. The non-zero momentum imparted by the photons causes an along-orbit acceleration on the asteroid changing its semi-major axis. In turn, the drift in semi-major axis drives the asteroids across a complex network of resonances with the planets of the Solar System and even with the major asteroids like Ceres \citep[][]{Morbidelli1999,Novakovic2015}. This forces the proper elements  $e$ and $i$ to change over time as well \citep[e.g.][]{Bottke2001,Broz2013a}. In fact, it is now clear by combining the HCM method with color and/or albedo information \citep[][]{Parker2008,Masiero2013} that most families are significantly more extended than previously thought, and are characterized by a broad ÒhaloÓ, which surrounds the ÒcoreÓ of the family. Only the core is detectable as a statistical significant asteroid clump in orbital elements space. This suggests that, over time, the core dissolves into the halo, so that families might become unrecognizable by the clustering method if they are old enough. This may explain the deficit of families older than 2 Gy, discussed above.
 
In order to attempt the identification of old families, in this paper we seek for a new method that does not rely on asteroid clumping in the space of proper eccentricity and inclination, the most dispersed parameters during long-term evolution. The idea is that, because the Yarkovsky effect is size-dependent (the semi-major axis drifts roughly at a speed proportional to $\frac{1}{D}$, where $D$ is the asteroid diameter, inwards or outwards depending on its retrograde/prograde spin), the families acquire a characteristic V-shape in the plane $a$ vs. $\frac{1}{D}$ or $a$ vs. $H$, where $H$ is asteroid absolute magnitude as seen for synthetic families in Figs.~\ref{fig.SynthsingleB}, \ref{fig.SynthsingleD} and \ref{fig.SynthsingleB_3_5} and the Erigone family in Fig.~\ref{fig.ErigoneD}. This shape is observed for all collisional families. Note that some families are only defined dynamically by a statistical clustering in ($a$,$e$,$i$) space that may not have a V-shape because they may not originate from a single collisional event, such as the Hertha family \citep[][]{Milani2014,Dykhuis2015}. However, the topic of this work focuses solely on single-collisional families. The V-shape is independent of the eccentricity and inclination distributions; mean motion resonances also have minimal effect on this shape, unless they deplete asteroids by pushing them out of the main belt, such as in the case of families bordering the powerful 7:2, 3:1, 5:2, 2:1 resonances with Jupiter \citep[][]{Walsh2013,Spoto2015}. For asteroid families located in mean motion resonances such as the Hilda and Eurybates families, the Yarkovsky drift in semi-major axis is transferred to the eccentricity \citep[][]{Broz2011,Milani2016}. Moreover, the width of the V-shape depends only on the Yarkovsky semi-major axis drift rate determined by asteroid densities, albedos, thermal inertia and rotation period of the asteroids \citep[][]{Vokrouhlicky1999a}, and on the age of the family \citep[][]{Vokrouhlicky2006b}.
 
Because of this last characteristic, the V-shape structures of asteroid families, previously identified with classical methods such as HCM, have been widely used over the last years to determine the age of known families \citep[e.g.][]{Nesvorny2003,Nesvorny2015a,Vokrouhlicky2006b,Bottke2007,Bottke2015a, Masiero2012, Walsh2013, Spoto2015}. The ÒV-shape methodÓ has also been used to identify family interlopers \citep[][]{Vokrouhlicky2006a,Nesvorny2015a}. These are bodies that are linked to the family by the HCM method, but fall outside of the V-shape structure of the family. In other words, interlopers are too far in semi-major axis from the family center to be part of the family, given their size. Finally, the analysis of the distribution of asteroids in the ($a$,$H$) plane and their physical properties, such as albedo and color, in the complex Nysa-Polana region, has allowed \citet{Walsh2013} to identify unambiguously two new families, Eulalia and new Polana , superseding the previous, confused family classification.
 
Expanding on the work of \citet{Walsh2013}, the goal of the paper is to make of the V-shape method a semi-automatic search tool, appropriate for finding old families severely dispersed in  $e$ and $i$, which cannot be identified by HCM. In Section~\ref{s.Vshapeformation}, we review the properties of the Yarkovsky evolution of asteroids, which define the V-shape, discussing also the consequences of the evolution of the spin axes of the asteroids, due to collisions. We also discuss the so-called Yarkovsky-O'Keefe-Radzievskii-Paddack (YORP) effect, a variation on the Yarkovsky effect that causes a torque on a small bodies and can change its rotation period and the direction of the spin axis \citep[][]{Rubincam2000} and stochastic YORP effects \citep[][]{Statler2009,Cotto-Figueroa2015}. In Section~\ref{s.Vshapeidentification} we describe two methods to identify that the asteroid distribution in the ($a$, $H$) or ($a$, $\frac{1}{D}$) planes has the prominent shape expected for a clump of asteroids spreading under the Yarkovsky effect, embedded in a dispersed background. We will test these methods in Section~\ref{s.Results}. First we will consider synthetic families (isolated and overlapping ones), in order to familiarize the reader on how the methods respond to the imprinted structures and on the appearance of the results. Then we will consider some known families, both young and old, showing how they could be blindly identified by the methods from an asteroid catalog. The conclusions and the perspectives to use these methods to identify currently unknown and old families will be discussed in Section~\ref{s.Discussion}.

\section{Family V-shape formation}
\label{s.Vshapeformation}
The initial velocity field contributes significantly to the semi-major axis spread of young and intermediate age families, such as the $\sim$ 280 Myr-old Erigone family, which a third of the spread in $a$ of its members is due to the initial velocity of the fragments \citep[][]{DellOro2004,Vokrouhlicky2006b}. The spread of fragments in $a$ due to initial velocity varies with fragments' $D$ as $ \left(\frac{1}{D}\right )^{\beta}$ \citep[][]{Cellino1999, Vokrouhlicky2006b, Michel2015}, where $\beta$ is assumed to be $\sim$1 following the observed range in fragment sizes in the Karin family \citep[][]{Nesvorny2002b}, causing smaller fragments to be more disperse than larger fragments \citep[][]{Cellino2009}. Their family members are spread in time by the Yarkovsky force from the center of the family in semi-major axis \citep[][]{Bottke2001,Bottke2006,Nesvorny2002a, Vokrouhlicky2015}. The effect of Yarkovsky force is orders of magnitude larger than the change in $a$ due to close encounters with massive asteroids on Gyr timescales for asteroids with D $<$ 20$\;$-$\;$40 km \citep[][]{Nesvorny2002a,Delisle2012,Carruba2013a} and asteroid collisions \citep[][]{DellOro2007}. The Yarkovsky force modifies the members' semi-major axes pushing the asteroids into secular and mean motion resonances (MMRs), which often modify the members' eccentricity and inclination by chaotic diffusion \citep[][]{Bottke2002, Carruba2005,Carruba2007,Carruba2011,Carruba2016a, Novakovic2015, Masiero2015b}. The powerful 7:3, 3:1, 5:2, 2:1 resonances with Jupiter amplify asteroids' eccentricities causing them to quickly evolve onto planet-crossing or sun-colliding orbits \citep[][]{Gladman1997,Farinella1998, Morbidelli1999, Bottke2002a}. Small resonances are also important as they can cause chaotic diffusion of family members' $e$ and $i$ as family members drift over them \citep[][]{Milani1994b,Nesvorny2002a}.

The semi-major axis drift rate, $\frac{da}{dt}$, caused by the Yarkovsky force  is proportional to $\frac{1}{D}$ and the cosine of the obliquity, $\phi$, creating a V-shape in $a$ vs. $\frac{1}{D}$ space with a border defined by a straight line \citep[][]{,Milani2014,Spoto2015}. Asteroid diameter can be converted into absolute magnitude, which transforms the straight lined V-shape in $a$ vs. $\frac{1}{D}$ space to a curved V-shape in $a$ vs. $H$ space \citep[][]{Nesvorny2003, Vokrouhlicky2006b}. Chaotic diffusion and secular resonances have little effect on the semi-major axes on Gyr timescales \citep[][]{Nesvorny2002a,Bottke2002a} preserving the V-shape on secular timescales.

In an ideal case, the family V-shape border would be traced by asteroids, which drifted in $a$ at the maximum rate allowed by their size over the full lifetime of the family, offset by the initial displacement caused by the original ejection velocity field as seen in the synthetic family V-shapes in Figs.~\ref{fig.SynthsingleB}-\ref{fig.SynthUniformD}. It must be noted that the drift rates for all asteroid family members in families older than 2 Gyr  have been globally affected by changing solar luminosity \citep[][]{Vokrouhlicky2006a,Carruba2015a}.

The magnitude of the Yarkovsky semi-major axis drift rate depends on the thermal inertia, asteroid diameter and rotation period \citep[][]{Bottke2006, Vokrouhlicky2015}. Nevertheless, we will use the $\frac{1}{D}$ functional form in the V-shape technique because we restrict our V-shape search to asteroids with $D$ between 1 and 40 km where the thermal inertial dependency on diameter is negligible and we assume a typical value for the rotation period \citep[][]{Delbo2015}.

The surface roughness of an asteroid also affects the magnitude of the Yarkovsky effect \citep[][]{Rozitis2012}. Recoil caused by thermal emission of photons off irregular macroscopic surface variations, such as regolith or small boulders, can dramatically increase the semi-major axis drift rate compared to an asteroid with a smooth surface.  Presently, there is no clear evidence showing a global surface roughness diameter dependence. 

The majority of asteroids in collisional families have slower than the maximum drift rate in $a$ due to rotation states that hinder the Yarkovsky effect, at least temporarily \citep[][]{Bottke2015a}. Yarkovsky semi-major axis drift rates are lower for asteroid obliquities far from 0 and 180 degrees and are almost non-existent for asteroid with extremely slow rotation rates \citep[][]{Vokrouhlicky2007, Vokrouhlicky2015}. The obliquities and rotation rates of asteroid families members are modified by the YORP effect \citep[][]{Vokrouhlicky2002, Capek2004}, spin-orbit resonances \citep[][]{Vokrouhlicky2003,Vokrouhlicky2006c} and the ``stochastic YORP" effect (see below) \citep[][]{Cotto-Figueroa2015,Bottke2015a}. Thus, in reality, the V-shape border is smeared because asteroids having the maximum drift rate over the full age of the family are rare. 

YORP can cause long-term variation of object's obliquity and rotation rate \citep[][]{Rubincam2000}. The end result of YORP is an object's obliquity reaching 0$\tdeg$, 180$\tdeg$ on $\sim$100 Myr timescales for km-sized objects \citep[][]{Vokrouhlicky2002,Capek2004, Scheeres2008}. This is comparable to axis reorientation time scales caused by non-catastrophic collisions \citep[][]{Farinella1998,Broz1999}. Meanwhile, objects either spin-up until they shed material or change shape due to rotational stress \citep[][]{Pravec2007,Pravec2010,Walsh2008} or spin-down until they reach a tumbling state during, which collisions can easily reset the spin \citep[][]{Vokrouhlicky2007, Breiter2015}. Both end states restart the YORP-driven evolution of the asteroid. The evolution between two resetting events is called a "YORP cycle".

YORP affects the semi-major axis dispersion of family fragments and depends on asteroid size and family age. Asteroids with D $\; > \;$ 5 km
 reach asymptotic obliquities due to YORP of 0$\tdeg$ and 180$\tdeg$ on much greater timescales than smaller asteroids because they preserve their initial obliquities on large time scales. Large asteroids will drift on average by a factor of $\frac{2}{\pi}$ less than the maximum rate expected for their size where $\frac{2}{\pi}$ is the average value of the cosine function used in the formula for the Yarkovsky drift rate of an asteroid (see section~\ref{s.Vshapeidentification} below). Small asteroids are more likely to drift at their maximal rate because they are quickly brought to an obliquity of 0$\tdeg$ and 180$\tdeg$ until their YORP cycle is reset.
 
The YORP effect is also dependent on the thermal conductivity of asteroids \citep[][]{Capek2004}. Different thermal conductivity values result in asteroids reaching asymptotic extreme obliquities (therefore maximum Yarkovsky semi-major axis drift rates) on different timescales. Obliquity values of 0$\tdeg$ and 180$\tdeg$ are easily reached by asteroids with thermal conductivity values typical for objects in the km size range  \citep[][]{Capek2004,Delbo2007}, which is supported by the observed obliquity distribution of km-sized asteroids \citep[][]{Hanus2011, Hanus2013b, Dykhuis2016}.

 In addition to the YORP cycles described above, small changes to the shape or surface features of asteroids caused by rotational stress or impacts \citep[][]{Walsh2012} can drastically change the strength of YORP \citep[][]{Statler2009}. Minute shape changes can cause an object's YORP evolution to reset stochastically \citep[][]{Cotto-Figueroa2015}. This ``Stochastic YORP" behavior primarily affects asteroids' spin rates and only has a minor effect on obliquity \citep[][]{Bottke2015a}. The rate at, which asteroids receive enough sub-catastrophic collisions to change their shape enough to modify the YORP evolution is a magnitude higher than that at, which the spin rate or axis are modified solely due to collisions \citep[][]{Farinella1998, Bottke2015a}.
 
A different V-shape function may be required for asteroids smaller than 5 km for families older than 2 Gyr as a consequence of the stochastic YORP effect \citep[][]{Bottke2015a}. The V-shape becomes more vertical at smaller objects sizes and for older family ages \citep[][see Figs. 13 and 15]{Bottke2015a}. The classical asteroid family V-shape described in  \citet{Vokrouhlicky2006b} may be preserved at larger asteroid sizes also for $>2$ Gyr-old families because the effects of stochastic YORP are less severe. The size frequency distribution, SFD, of $>2$ Gyr-old families is typically shallow so that the family is dominated by $D$ $\gtrsim$ 3 km asteroids. This is due to the dynamical and collisional erosion of the $D$ $\lesssim$ 3 km asteroid population on Gyr timescales while $D$ $\gtrsim$ 3 km asteroids remain preserved \citep[][]{Marchi2006,Carruba2015a, Carruba2016b}. Thus, when searching for very old families we may neglect the stochastic YORP effect and look for families by searching for a classic V-shape. 

 Additional spin properties affect the Yarkovsky-driven drift rate of asteroids and make the family structure deviate relative to it's ideal V-shape. The spin state distribution of asteroids can be non-uniform as a result of overlapping spin-orbit resonances \citep[][]{Vokrouhlicky2003,Vokrouhlicky2006c}. Obliquity clustering of asteroids located in central belt families is caused by secular resonances between the asteroid obliquity precession rate and the precession rate of Saturn's longitude of node \citep[][]{Slivan2002,Slivan2003,Vokrouhlicky2003,Vokrouhlicky2006c}. Similar obliquity clustering may also be present among inner main belt  Massalia and Flora asteroid family members \citep[][]{Vrastil2015, Dykhuis2016}. Objects with obliquities locked in spin-orbit resonances have a reduced Yarkovsky semi-major axis drift compared to the one they would have if their obliquities were either $0\tdeg$ or $180\tdeg$. Thus, the asteroids locked in spin-orbit resonances will exhibit a deficit of migration relative to other unlocked asteroids of comparable size. In fact, unlocked asteroids have a displacements that is a result of drift rates governed by their size and more uniformly distributed obliquities over the age of the asteroid family. The $a$ vs. $H$, $\frac{1}{D}$ distribution of asteroid families may then be a combination of V-shapes caused by overlapping populations of spin-orbit resonance locked asteroids and unlocked asteroids.
 
Additional effects can change the placement of asteroids relative to the nominal V-shape structure. Asteroid family members may be offset in semi-major axis due to close encounters with massive asteroids \citep[][]{Nesvorny2002a,Carruba2003,Carruba2013a,Delisle2012}. For the largest asteroids, i.e., the largest family remnants, the effect of encounters can dominate over the  Yarkovsky semi-major axis drift and place these objects outside of the V-shape \citep[][]{Walsh2013, Spoto2015}.

\section{Family V-shape identification}
\label{s.Vshapeidentification}
Family V-shapes are used to measure the age of families (\eg~\citet{Nesvorny2003,Nesvorny2015a}, \cite{Vokrouhlicky2006b}, \citet{Bottke2007,Bottke2015a}, \citet{Masiero2012}, \citet{Walsh2013} and \citet{Spoto2015}) using estimates from a linear Yarkovsky semi-major axis drift models \citep[][]{Vokrouhlicky1999a}. However, \citet{Walsh2013} firstly used the V-shape to the particular case of identify the families of Eulalia and new Polana . Here we expand on the work of Walsh et al. and we develop further the method to make it a general technique can to find collisional asteroid families.

As we explained above, asteroid families, whose members' proper elements $e$ and $i$ have become too dispersed due to chaotic diffusion can be identified by searching for correlations in $a$ vs. $\frac{1}{D}$, $H$ space. The size-dependent Yarkovsky force gives a family the V-shape in $a$ vs. $\frac{1}{D}$,$H$ distribution on Myr time-scales. In practice, it is possible for a family to obtain a V-shape on shorter timescales due to the contribution of the initial velocity field.

The sides of the V-shape in $a$ vs. $\frac{1}{D}$ space is
\begin{equation}
\label{eqn.Davsdadt}
a - a_c \; = \; \frac{da}{dt}(D) \; \Delta t
\end{equation}
where $a_c$ is the family center,  $\frac{da}{dt}(D)$ is the size dependent maximal Yarkovsky semi-major axis drift rate and $\Delta t$ is the age of the family. The drift rate can be recalculated for different bulk and surface densities, orbit, rotation period, obliquity and thermal properties \citep[][]{Bottke2006, Chesley2014, Spoto2015}. We define the drift rate $\frac{da}{dt}(D)$ as

\begin{equation}
\label{eqn.yarkorate}
\frac{da}{dt} (D) \; = \; \left ( \frac{da}{dt} \right )_0 \; \left ( \frac{1329 \km}{D} \right ) \; \left ( \frac{1}{\rho} \right ) \; \left ( \frac{\au}{\myr} \right ) \left ( \frac{1 - A}{1 - A_{0}} \right )
\end{equation} 
from \citep[][]{Walsh2013}. The Yarkovsky drift rate, $\left ( \frac{da}{dt} \right )_0$ is $\sim 2.8 \x 10^{-7}$ au Myr$^{-1}$, the Yarkovsky semi-major axis drift rate for a 1329 km asteroid in the inner Belt with a density, $\rho$, of 1.0 $\gpcmc$, thermal conductivity $\cal{K}$ $\sim$ 0.01 - 0.001 W m$^{-1}$ K$^{-1}$, Bond albedo, $A_0$, of  0.02 \citep[][]{Harris2002,Spoto2015}, rotation period 3.5 h and obliquity 60$\tdeg$. Notice that the fastest drifting asteroids have obliquity equal to 0$\tdeg$ and therefore they drift at twice the speed reported above.  However, asteroids that drifted at maximum speed over the entire family age are probably rare and difficult to identify relative to the background. Therefore we expect that the average drift rate for obliquity 60 deg is a more appropriate number to use.

The width of the V-shape in $a$ vs. $1/D$ space can be defined by the constant $C$

\begin{equation}
\label{eqn.Cpvdadtvsdeltat}
C \; = \; \Delta t \left ( \sqrt \pv \; \left ( \frac{da}{dt} \right )_0 \right )
\end{equation}
where $\pv$ is the visual albedo, which is assumed to be the same for all family members (an assumption well supported by observations; \citep[][]{Masiero2013}. Typical $\pv$ values of 0.05 and 0.15 are used for C- and S-type asteroids, respectively \citep[][]{Masiero2011, Masiero2015a}

Combining Eqs.~\ref{eqn.Davsdadt}, \ref{eqn.yarkorate} and \ref{eqn.Cpvdadtvsdeltat} we define the border of the V-shape in reciprocal diameter, $\frac{1}{D}$ or $D_r$, space as
\begin{equation}
\label{eqn.apvDvsC}
D_r(a,a_c,C,\pv) \; = \; \frac{\left | a - a_c \right | \; \sqrt{\pv} }{1329 \km \; C}
\end{equation}

Defining diameter, $D$, as $D = 2.99 \x 10^8 \; \frac{10^{0.2 \; (m_\odot  \; - \; H)}}{\sqrt{\pv}}$ \citep[][]{Bowell1988}, where $m_\odot \; = \; -26.76$ \citep[][]{Pravec2007}. The border of the V-shape in absolute magnitude, $H$, space is 
\begin{equation}
\label{eqn.aHvsC}
H(a,a_c,C) \; = \; 5 \; \mathrm{log}_{10}\left ( \frac{\left | a - a_c \right |}{C} \right )
\end{equation}
\citep[see also][]{Vokrouhlicky2006b}. Different physical properties will not change the functional forms of Eqs.~\ref{eqn.apvDvsC} and \ref{eqn.aHvsC} and will only change the calculated age for a given $C$ from Eq.~\ref{eqn.Cpvdadtvsdeltat}, provided these properties are not size dependent as discussed in Section~\ref{s.Vshapeformation}.

The V-shape technique is limited to asteroids with an upper limit of $H\;$$<\;$16 or $D\;$$\gtrsim \;$3 km assuming a $\pv \; = \; 0.1$ because stochastic YORP may cause the portion of the V-shape defined by smaller asteroids to have a larger slope compared to Eq.~\ref{eqn.aHvsC} as described in Section~\ref{s.Vshapeformation}. The distortion of the V-shape caused by stochastic YORP is enhanced on $>\;2$ Gyr timescales and starts to affect the border defined by larger objects. 

The absolute magnitude range of asteroids used when applying this technique has a lower limit of $H$ $>$ 12 or $D$$\;\lesssim\;$20 km assuming $\pv = 0.1$ because larger family members should not have been affected by the Yarkovsky semi-major axis drift significantly and they could potentially be displaced relative to the nominal V-shape by the effects discussed in the previous section (e.g., close encounters with massive asteroids). Displacement in $a$ due to Yarkovsky semi-major axis drift for 20 km body is only $\sim$0.01 au over 1 Gyr. Using $H \;> \;12$ as a lower bound also prevents larger asteroids that may be interlopers and are not alined with the V-shape from being inlcuded in the V-shape search.

It is clear that the search for a V-shape can be done equivalently either in the $a$ vs. $\frac{1}{D}$ plane, looking for a border with the functional form defined in Eq.~\ref{eqn.apvDvsC}, or in the $a$ vs. $H$ plane, using the functional form Eq.~\ref{eqn.aHvsC}. The choice between using $\frac{1}{D}$ or $H$ depends on the asteroid catalog. Moreover, as we will see in Section~\ref{s.Results}, before a blind search for families is done, it is crucial to select the asteroids that have uniform physical properties by restricting to a range in albedos. If the albedos are used, then the asteroid catalog necessarily has the $D$ measurements to use in the V-shape search. 

A search in the $a$ vs. $H$ plane is preferred if the catalogue used contains more accurately calibrated $H$ measurements compared to the measurements in the MPC catalogue \citep[such as those in][]{Veres2015}. Here, asteroids must be selected by their albedos, so only asteroids in the improved $H$ catalogue, which also have $D$ measurements are used. The number of asteroids with improved $H$ measurements from \citet{Veres2015} is less than the number of asteroids with $D$ measurements by $\sim$~30$\%$. Therefore, the advantage of using the V-shape search in $a$ vs. $H$ over $a$ vs. $D$ depends on the local abundance of improved H magnitudes at the location of the search in the main belt. 

Below, we will explain two methods for identifying a V-shape border. For sake of example, we will present the first in $a$ vs. $\frac{1}{D}$ and the second in $a$ vs. $H$ but each method can be used in both coordinate planes. 

\subsection{Border method}
\label{s.border}

\citet{Walsh2013} found that the borders of the V-shapes of the Eulalia and new Polana  family could be identified by the peak in the ratio $\frac{N_{in}}{N_{out}}$ where $N_{in}$ and  ${N_{out}}$ are the number of asteroids falling between the curves defined by Eq.~\ref{eqn.apvDvsC} for values $C$ and $C_-$ and  $C$ and $C_+$, respectively, with $C_-=C-dC$ and $C_+=C+dC$, namely:

\begin{equation} 
\label{eq.border_method_N_outer}
  N_{out}(a_c,C,dC)
    = \; \Sigma_j 
       \; w(D_j) \;
       \int\limits_{a_1}^{a_2} da
    	\int\limits_{D_r(a, a_c,C_+, \pv)}^{D_r(a, a_c,C, \pv)} dD_r
       \; \delta(a_{j}-a)
       \; \delta( D_{r,j}-D_r )
\end{equation}

\begin{equation} 
\label{eq.border_method_N_inner}
  N_{in}(a_c,C,dC)
    = \; \Sigma_j 
       \; w(D_j) \;
       \int\limits_{a_1}^{a_2} da
    	\int\limits_{D_r(a, a_c,C, \pv)}^{D_r(a, a_c,C_-, \pv)} dD_r
       \; \delta(a_{j}-a)
       \; \delta( D_{r,j}-D_r )
\end{equation}

The symbol $\Sigma_j$ indicates summation on the asteroids of the catalog, with semi-major axis $a_j$ and reciprocal diameter $D_{r,j}$. The symbol $\delta$ indicates Dirac's function, and $a_1$ and $a_2$ are the low and high semi-major axis range in which the asteroid catalog is considered.  The function $w(D)$ weighs the right-side portions of Eqs.~\ref{eq.border_method_N_outer} and \ref{eq.border_method_N_inner} by their size so that the location of the V-shape in $a$ vs. $D_r$ space will be weighted towards its larger members. We use $w(D) = D^{2.5}$, in agreement with the cumulative size distribution of collisionally relaxed populations and with the observed distribution for MBAs in the $H$ range $12\; < \;H \;< \;16$ \citep[][]{Jedicke2002}. Asteroids in families whose parent body has undergone catastrophic disruption have an SFD slope similar to the SFD slope of background asteroids in the main belt due to collisional  evolution of their family members over Myr timescales \citep[][]{Morbidelli2003b}. Different SFD slopes could be used in principle for asteroid families resulting from different kinds of disruption events \citep[e.g.,][]{Tanga1999,Bottke2005a} but this is beyond the scope of the current study.

The value of $dC$ is an arbitrary value. It can be much smaller, to within a few 10$\%$ of the family V-shape's $C$ value if the number density of asteroids on a V-shape's border is high and the border has a clear edge. The ratio of $N_{in}$ to $N_{out}$ will be high enough to identify the family with a small value of $dC$ if there is a steep drop in the number of asteroids outside of the border.  A larger value of $dC$ up to 40$\sim$50$\%$ of the family V-shape's $C$ value is needed if the V-shape border is diffuse and has a lower number density. The inner and outer V-shapes must be wide enough to include enough asteroids in the inner V-shape and measure a $N_{in}$ to $N_{out}$ ratio high enough to identify the family V-shape. The V-shape can include interlopers or asteroids which are not apart of the family V-shape if value of $dC$ is used that is too large. A peak value in $\frac{N_{out}(a_c,C,dC)}{N_{in}(a_c,C,dC)}$ (top panel of Fig.~\ref{fig.SynthsingleB}) indicates the best fitting values of $a_c$ and $C$ for a family V-shape using Eq.~\ref{eqn.apvDvsC} (bottom panel of Fig.~\ref{fig.SynthsingleB}). A peak in $\frac{N_{out}(a_c,C,dC)}{N_{in}(a_c,C,dC)}$ is significant if it is significantly greater than 2 and statistically significant compared to the surrounding values of $\frac{N_{out}(a_c,C,dC)}{N_{in}(a_c,C,dC)}$ in $a_c$ vs. $C$ space. The number for $\frac{N_{out}(a_c,C,dC)}{N_{in}(a_c,C,dC)}$ for a family's V-shape determined to be statistically significant must be considered separately each family V-shape in the case of overlapping or nearby families in $a$ vs. $D_r,H$ space

\subsection{Density method}
\label{s.density}

Another method to identify the characteristic V-shape of a family is to look for the region of maximal asteroid density $\rho$. We define $\rho$ as:

\begin{equation} 
\label{eq.density_method}
  \rho(a_c,C,dC,\pv)
    = \frac{\; \Sigma_j 
       \; w(D_j) \;
       \int\limits_{a_1}^{a_2} da
    	\int\limits_{D_r(a, a_c,C, \pv)}^{D_r(a, a_c,C_-, \pv)} dD_r
       \; \delta(a_{j}-a)
       \; \delta( D_{r,j}-D_r )}{\int\limits_{a_1}^{a_2}  da
    	\int\limits_{D_r(a, a_c,C,\pv)}^{D_r(a, a_c,C_-,\pv)} \; dD_r}
\end{equation}

Peaks in $\rho$ indicate the best fit for $a_c$ and $C$ in Eq.~\ref{eqn.aHvsC} (top panel of Fig.~\ref{fig.SynthsingleD}). Similar to the border method, smaller $dC$ values are favored for higher asteroid densities and lower densities larger values of $dC$.

\subsection{Comparison with known families}
\label{S.Ages}

In Section~\ref{s.Results}, the identification of known families with the V-shape method can be cross checked with previous results by comparing C values measured from the V-shape identification to the published values of the C parameters \citep[][]{Dykhuis2014, Nesvorny2015a}. These authors also used the V-shapes to determine the ages of the families; however, they determine the slopes of the V-shapes with different techniques. The purpose of the comparison is to verify whether the optimal V-shapes we find with our methods are consistent with theirs. The age of the family is typically calculated using the value of $C$ determined by the V-shape search method and the approximate drift rate $(\frac{da}{dt})$ determined from Eq.~\ref{eqn.Cpvdadtvsdeltat}

\begin{equation}
\label{eq.fam_age}
t_{age} \; = \; \frac{C}{\left ( \sqrt \pv \; \left ( \frac{da}{dt} \right ) \right )}
\end{equation}

The age determined by Eq.~\ref{eq.fam_age} is an upper limit on the family age because the value of $C$ has to be corrected to account for the initial ejection velocity field. The typical magnitude of the initial ejection velocity is typically correlated to the escape speed from the parent body \citep[][]{Vokrouhlicky2006b,Walsh2013,Nesvorny2015a}. The latter can be estimated once the family members are identified and the mass of the parent body is determined from the sum of the masses of the family members after correcting for observational selection effects, dynamical and collisional depletion of the family members over the age of the family. Determining family ages and parent body size is beyond the scope of this work, so the age determined by using $C$ found with the V-shape method and Eq.~\ref{eq.fam_age} will be used as an approximate comparison with known ages of synthetic or real families.

The best fit value of $C$ determined by the density method is systematically lower than the value of $C$ determined by the border method because the border method is sensitive to the location of the ``front runner asteroids" (those who drifted at the maximal rate) whereas the density method is sensitive to the location of the bulk of the family population. Because of all the reasons explained in Section~\ref{s.Vshapeformation}, the bulk of the family population has drifted less than the front runner asteroids. The differences in $C$ between the border and density methods is exacerbated by physical effects $\eg$ of stochastic YORP \citep[][]{Bottke2015a} and possible Slivan states such as asteroids in the cases of the Flora \citep[][]{Dykhuis2015, Dykhuis2016} and Koronis families \citep[][]{Vokrouhlicky2003}. Thus the value of $C$ for a family's V-shape, determined by the density method, should be used as a lower limit for family age computations because the method will be more weighted towards the density enhancement away from the actual V-shape border as it is shown in Figs.~\ref{fig.SynthsingleB} and \ref{fig.SynthsingleD}.

\section{Results}
\label{s.Results}

\subsection{Test of the methods on synthetic families}

The V-shape detection method is tested on synthetic asteroid families with definitive and unconfused V-shapes. Synthetic families were created by simulating the dispersal of family fragments following a catastrophic disruption assuming an size-dependent ejection velocity field \citep[][]{Zappala2002}. The size distribution of synthetic family fragments were scaled from the asteroid family fragment SFD model of \cite{Durda2007}, where the mass of the second largest remnant is a free variable, but the size distribution of remaining fragments is propagated to smaller sizes starting with the second largest fragment, using an incremental SFD with a slope of 2.85 \citep[see ][]{Leinhardt2012}. The fragments are evolved in $a$ vs. $\frac{1}{D}$, $H$ space using the Yarkovsky model of \citet{Vokrouhlicky2006b} and the cube root of the sine of the obliquity distribution of particles used to weight the distribution towards $0\tdeg$ and $180\tdeg$ to simulate the long-term effects of YORP obliquity evolution. The values of $C$ found with the V-shape method are lower limits due in part because obliquities of the asteroids are assumed to remain constant throughout the age of the synthetic family, and stochastic YORP, YORP cycling of fragments' obliquity values are not modeled. 

\subsubsection{Single V-shape family}
\label{s.SingleV}

\begin{figure}
\centering
\hspace*{-2.7cm}
\vspace{-0.5 cm}
\ifincludeplots
\ifgrayscale
\includegraphics[scale=0.50]{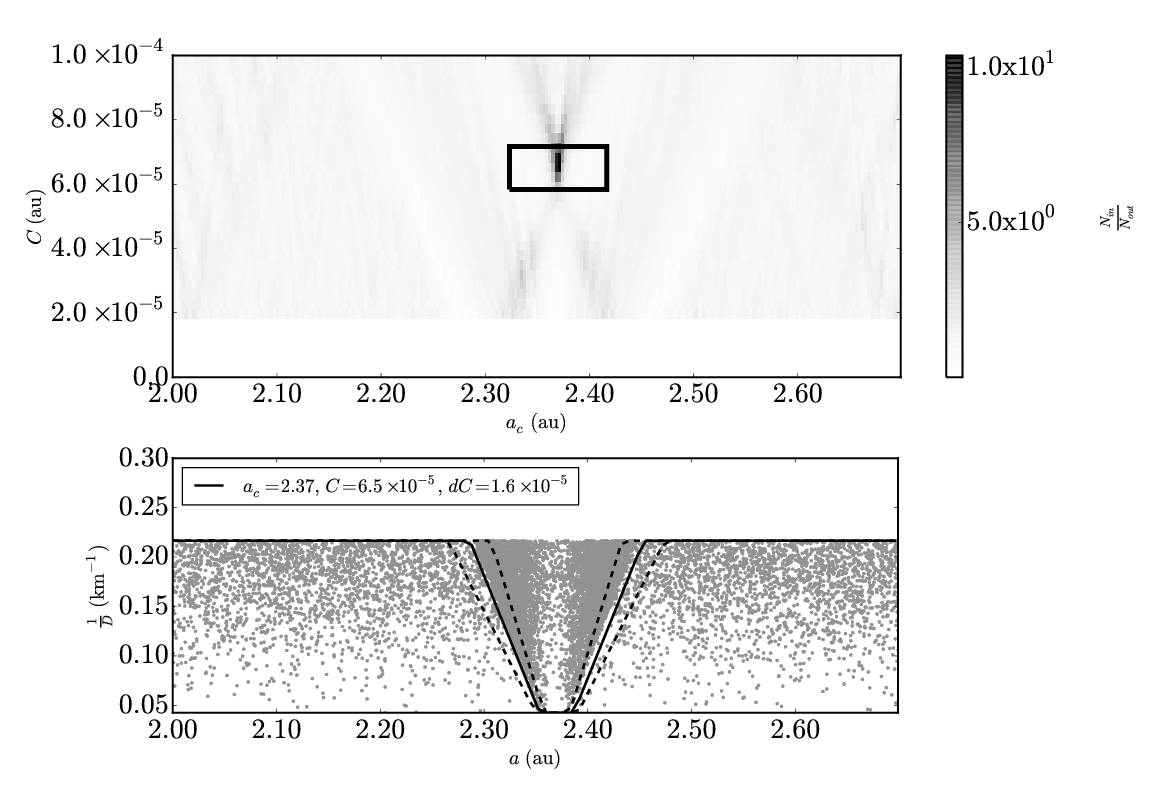}
\else
\includegraphics[scale=0.50]{Synth_single_border.png}
\fi
\else
I am not enabling plots.
\fi
\caption{Application of the border method. (Top panel) The ratio between the number of asteroids in the outer V-shape to the number of asteroids in the inner V-shape in the $a_c$-$C$ range, ($a_c\pm \frac{\Delta a_c}{2}$,$C\pm \frac{\Delta C}{2}$) where $\Delta a_c$ is equal to $3.0 \times 10^{-3}$ au and $\Delta C$, not to be confused with $dC$, is equal to $3.0 \times 10^{-6}$ au, for a single synthetic family. The box marks the peak value in $\frac{N_{out}(a_c,C,dC)}{N_{in}(a_c,C,dC)}$ for the synthetic family V-shape. (Bottom Panel) $D_r(a,a_c,C,\pv)$ is plotted for the peak values with the primary V-shape as a solid line where $\pv = 0.05$. The dashed lines mark the boundaries for the area in $a$ vs. $D_r$ space for $N_{in}$ and $N_{out}$ using Eq.~\ref{eqn.apvDvsC},  $D_r(a,a_c,C\pm dC,\pv)$ where $dC \; = \; 1.6 \x 10^{-5}$ au. The X-shaped region in the top panel represents values of $a_c$ and $C$ resulting in elevated values of $\frac{N_{out}(a_c,C,dC)}{N_{in}(a_c,C,dC)}$ because the inner and outer V-shapes partially cover the family V-shape. A peak value of $\frac{N_{out}(a_c,C,dC)}{N_{in}(a_c,C,dC)}$ occurs at the center of the X-shape when the inner and outer V-shapes fully contain the family V-shape.}
\label{fig.SynthsingleB}
\end{figure}

\begin{figure}
\centering
\hspace*{-2.7cm}
\ifincludeplots
\ifgrayscale
\includegraphics[scale=0.50]{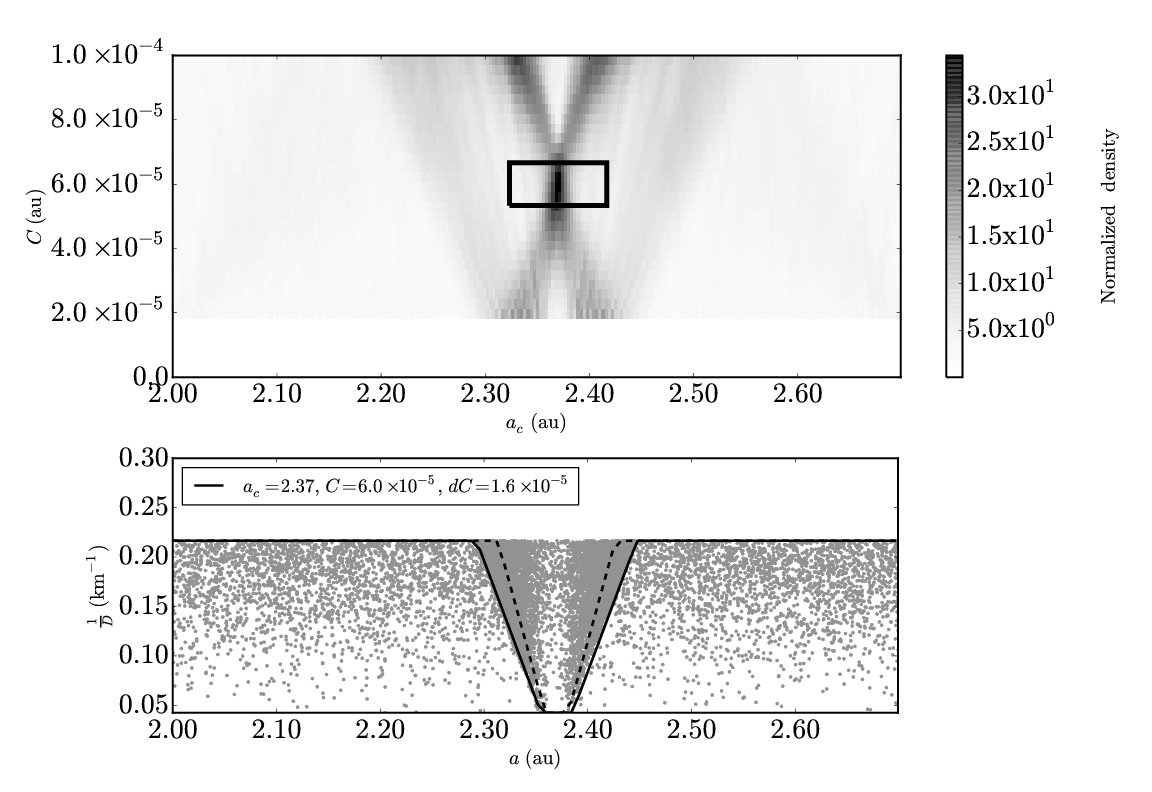}
\else
\includegraphics[scale=0.50]{Synth_single_density.png}
\fi
\else
I am not enabling plots.
\fi
\caption{Application of the density method. (Top panel) The normalized density in units of km au$^{-1}$ for asteroids in the inner V-shape in the $a_c$-$C$ range, ($a_c\pm \frac{\Delta a_c}{2}$,$C\pm \frac{\Delta C}{2}$) where $\Delta a_c \; = \; 3.0 \times 10^{-3}$ au and $\Delta C \; = \; 3.0 \times 10^{-6}$ au for a single synthetic family. The box marks the peak value in the normalized density for the synthetic family V-shape. (Bottom Panel) $D_r(a,a_c,C,\pv)$ is plotted for the peak values with the primary V-shape as a solid line where $\pv = 0.05$. The dashed line mark the boundary for the area in $a$ vs. $D_r$ space for the inner V-shape using Eq.~\ref{eqn.apvDvsC},  $D(a,a_c,C - dC,\pv)$ where $dC \; = \; 1.6 \x 10^{-5}$ au. The X-shaped region in the top panel represents elevated values of $\rho$ caused by when the central and inner V-shapes partially contain the family V-shape. A peak value of $\rho$ occurs when the central and inner V-shapes fully contain the family V-shape.}
\label{fig.SynthsingleD}
\end{figure}

\begin{figure}
\centering
\hspace*{-2.7cm}
\ifincludeplots
\ifgrayscale
\includegraphics[scale=0.50]{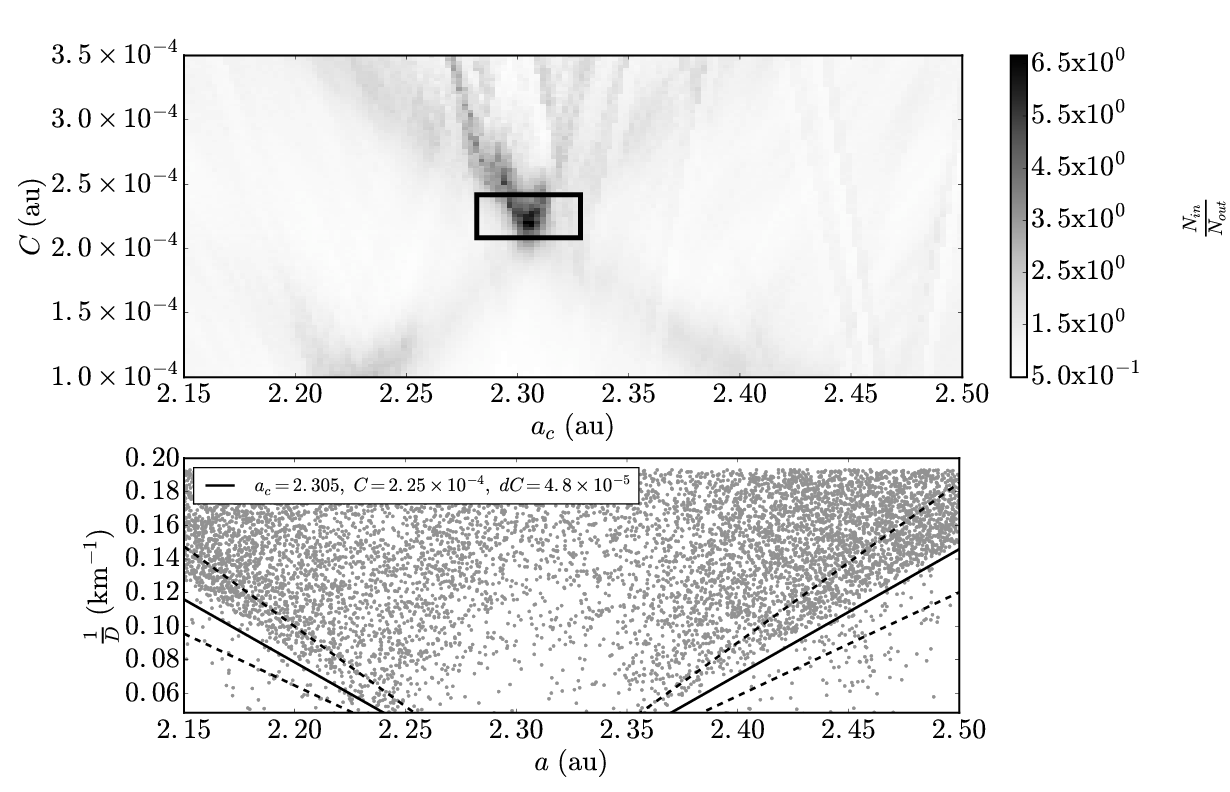}
\else
\includegraphics[scale=0.50]{Synth_single_border_grey_3_5_gyr.png}
\fi
\else
I am not enabling plots.
\fi
\caption{Application of the border method to a 3.5 Gyr-old synthetic family. (Top panel) The ratio between the number of asteroids in the outer V-shape to the number of asteroids in the inner V-shape in the $a_c$-$C$ range, ($a_c\pm \frac{\Delta a_c}{2}$,$C\pm \frac{\Delta C}{2}$) where $\Delta a_c \; = \; 2.5 \times 10^{-3}$ au and $\Delta C \; = \; 2.5 \times 10^{-6}$ au for a single synthetic family. The box marks the peak value in the normalized density for the synthetic family V-shape. (Bottom Panel) $D_r(a,a_c,C,\pv)$ is plotted for the peak value at $a_c \; = \; 2.305 \; \au$ and $C \; = \; 2.25 \times 10^{-5} \; \au$ with the primary V-shape as a solid line where $\pv = 0.05$. The dashed line mark the boundary for the area in $a$ vs. $D_r$ space for the inner V-shape using Eq.~\ref{eqn.apvDvsC},  $D_r(a,a_c,C\pm dC,\pv)$ where $dC \; = \; 4.8 \x 10^{-5}$ au.}
\label{fig.SynthsingleB_3_5}
\end{figure}

A synthetic family modeled after the C-type Erigone family was generated at $\left (a ,e , \sin i \right ) \; = \; \left (2.37, 0.21, 0.08 \right )$ using 50,000 particles generated from a SFD with a slope of 2.85 for asteroids with $4.5 \km \;\lesssim \; D \; \lesssim \; 50.0 \km$ (where the second largest fragment was $\sim$50 km, see bottom panel of Fig~\ref{fig.SynthsingleB}, zoomed to $0.04\;\invkm \;\lesssim \; D_r \; \lesssim \; 0.30\; \invkm$). Particles were assumed to have density and $\pv$ of 1.0 $\gpcmc$ and 0.05 respectively, typical values for C-types \citep[][]{Yeomans1997,Marchis2008}. The parent body of the asteroid family has a diameter of 160 km an escape speed of $\sim$ 60 $\mps$. An additional $10$ $\mps$ ejection speed were given to the fragments and uniformly distributed with respect to the radial, transverse and normal velocity components. The escape speed and additional ejection speed correspond to a maximum initial $a$ displacement of $\sim 1.4 \times 10^{-5}$ au for a 5 km diameter asteroid. The eccentricity and inclination distributions were determined using Gaussian scaling \citep[described in ][]{Zappala2002}, although the dispersion of fragments' eccentricities and inclinations were scaled up by 2x and 3x respectively to obtain a better qualitative match to the structure of the Erigone family when a similar sized synthetic family was dispersed by $\sim$280 Myr. The synthetic family members' semi-major axes were evolved for $\sim$800 Myr and removed from the simulation based on the size-dependent disruption timescale in \citet{Farinella1998}. At the end of the simulation, $\sim$6,000 particles remained with the majority being removed from the simulation due to collisional evolution and observational selection effects modeled after the Wide-field Infrared Survey Explorer (WISE) survey \citep[][]{Masiero2011}. The remaining asteroids were placed in a background of 6,000 particles randomly distributed with a uniform distribution in $a$ and with $4.5 \km \;\lesssim \; D \; \lesssim \; 50.0 \km$ using a SFD with a slope of 2.85.

The border method was applied to the single synthetic family using asteroids with 2.0 au $< \; a \;<$ 2.7 au and $4.5 \km \;\lesssim \; D \; \lesssim \; 25.0 \km$. Eqs.~\ref{eq.border_method_N_outer} and \ref{eq.border_method_N_inner} are integrated using the interval ($-\infty$,$\infty$) for the Dirac delta function $\delta(a_{j}-a)$ and the interval [$0.04\;\invkm, 0.22\; \invkm$] for the Dirac delta function $\delta( D_{r,j}-D_r )$. Eq.~\ref{eqn.apvDvsC} is truncated to $0.04 \; \invkm$ for $D_r \lesssim 0.04 \; \invkm$ and to $0.22 \; \invkm$ for $D_r \gtrsim 0.22 \; \invkm$. The peak in $\frac{N_{in}}{N_{out}}$ at $(a_c, C) \; = \; (2.37 \; \mathrm{au}, \;6.5 \times 10^{-5} \; \mathrm{au})$ (Fig.~\ref{fig.SynthsingleB}, top panel) corresponds to the location of the family's V-shape in the bottom panel of Fig.~\ref{fig.SynthsingleB}. The peak value of $\frac{N_{in}}{N_{out}}$ is 11.8, $\sim$ 22 standard deviations above the mean of 1.1 for $\frac{N_{in}}{N_{out}}$ in the range 2.0 au $< \; a \;<$ 2.7 au and 1.8 $\times \; 10^{-5}$ au $< \; C \;<$ 1.0 $\times \; 10^{-4}$ au. The solid line in the bottom panel represents the nominal V-shape and dashed lines representing V-shapes for the inner and outer borders described by Eqs.~\ref{eq.border_method_N_outer} and \ref{eq.border_method_N_inner}. This value of $C$ corresponds to an age of $\sim$1 Gyr. Revising the value of $C$ for the initial displacement of the fragments by subtracting the maximum initial semi-major axis displacement of $\sim 1.4 \times 10^{-5}$ results in an age of $\sim$800 Myr matching the duration time of the simulation.

The density method finds an identical value of $a_c$ of 2.37 and a $\sim10\%$ lower value for $C$ of $6.0 \times 10^{-5}$ (top panel of Fig.~\ref{fig.SynthsingleD}) compared to the result from the border method. The peak value of $\rho$ is 34.2, $\sim$ 6 standard deviations above the mean of 5.9 for $\rho$ in the range 2.0 au $< \; a \;<$ 2.7 au and 1.8 $\times \; 10^{-5}$ au $< \; C \;<$ 1.0 $\times \; 10^{-4}$ au. The density method finds systematically lower values for $C$ as described in Section~\ref{S.Ages}, resulting in a younger age of 720 Myr compared to the 800 Myr age calculated from the value of $C$ found with the border method. The peak in $(a_c, C)$ is also larger in the density method compared to the peak found with the border method and more elongated in $C$ because the density of asteroids in the synthetic family V-shape is relatively constant in the area just before the edges of the V-shape resulting in similar density values over a range of $C$ values representing V-shapes of different widths.

The border method was applied to an older version of the single synthetic family generated at  $\left (a ,e , \sin i \right ) \; = \; \left (2.305, 0.21, 0.08 \right )$ where its member's semi-major axes were time evolved for 3.5 Gyrs. The synthetic family members were imbedded in the real inner main belt population with orbital elements between $2.15 \au \; < \; a \; < \; 2.50 \au$, $0.0 \; < \; e \; < \; 0.2$, $0.0 \; < \; \sin i \; < \; 0.12$ and $\pv$ between $0.1 \; < \; \pv \; < \; 0.3$ and $H$ between $10.0 \; < \; H \; < \; 15.3$. The lower limit on $H$ of 12.0 was chosen to limit the technique to being used on asteroids with $D \; \sim$20 km, assuming a $\pv \; = \; 0.05$, or smaller, because asteroids with 20 km diameter or smaller are significantly affected by the Yarkovsky effect on Gyr-time scales \citep[][]{Bottke2006, Delisle2012}. The upper limit on $H$ of 15.3 was chosen because asteroids with $D \; \sim$5 km, assuming a $\pv \; = \; 0.05$, have been shown to survive the last 3.8 Gyrs of dynamical evolution \cite[][]{Marchi2006,Carruba2015a,Carruba2016c}. A peak in $\frac{N_{in}}{N_{out}}$ was found at $a_c \; = \;$ 2.305 au and $C \; = \; 2.25 \times 10^{-4}$ au. The peak value of $\frac{N_{in}}{N_{out}}$ is more than 9 standard deviations above the mean value for $\frac{N_{in}}{N_{out}}$ in the range 2.15 au $< \; a \;<$ 2.5 au and 1.0 $\times \; 10^{-4}$ au $< \; C \;<$ 3.5 $\times \; 10^{-4}$ au. A similar result was found with the density method applied to the 3.5 Gyr-old synthetic family. The value of $C \; = \; 2.25 \times 10^{-4}$ found with the border method corresponds to an age of $\sim$3.5 Gyrs. The value of $C \; = \; 2.25 \times 10^{-4}$ for a synthetic 3.5 Gyr-old family may be a lower limit on the $C$ value of a real 3.5 Gyr-old family's V-shape because the simulation producing the synthetic family does not include effects such as stochastic YORP. The inclusion of stochastic YORP in the simulation may cause the value of $C$ to increase significantly for families with ages on Gyr time compared to synthetic family V-shapes simulated with static YORP. The difference in $C$ values between family V-shapes generated with and without stochastic YORP is exacerbated for families with Gyr ages compared to younger family V-shapes with younger ages generated with and without stochastic YORP \citep[see section 5 of][]{Bottke2015a}.

\subsubsection{Half V-shape family}
\label{s.HalfV}

Several asteroid families are located near powerful MMRs with Jupiter and have their V-shape sculpted in $a$ vs. $\frac{1}{D}$ space into a half V-shape at the location of the resonance. Examples include the new Polana and Eulalia families crossed by the 3:1 MMR with Jupiter (bottom panel of Figs.~\ref{fig.NewPolanaB} and \ref{fig.EulaliaD}). To test the capabilities of the V-shape techniques to detect families with half V-shapes, a synthetic family is generated at $\left (a ,e , \sin i \right ) \; = \; \left (2.49, 0.21, 0.08 \right )$ with 50,000 particles generated from a SFD with a slope of 2.85 for asteroids with $4.5 \km \;\lesssim \; D \; \lesssim \; 50.0 \km$ (where the second largest fragment was $\sim$50 km, see bottom panel of Fig.~\ref{fig.SynthhalfB}) using the same synthetic family generation technique from Section~\ref{s.SingleV}. The fragments semi-major axes were evolved over 800 Myr and the effect of the 3:1 MMR on the family V-shape was approximated by removing asteroids with when their semi-major axis exceeded 2.49 au in addition to removing particles due to collisional evoltion. About 3,000 particles remained at the end of the simulation with the majority removed due to collisional evolution, crossing into the 3:1 MMR or occluded due to observational selection effects. The remaining asteroids were placed in a background of 6,000 particles uniformly distributed in $a$ and with $4.5 \km \;\lesssim \; D \; \lesssim \; 50.0 \km$ using a SFD with a slope of 2.85. Particles were assumed to have density and $\pv$ of 1.0 $\gpcmc$ and 0.05 respectively. The parent body of the asteroid family has a diameter of 160 km and an escape speed of $\sim$ 60 $\mps$. An additional $10$ $\mps$ ejection speed were given to the fragments and uniformly distributed with respect to the radial, transverse and normal velocity components.

The border method was applied to asteroids with 2.39 au $< \; a \;<$ 2.49 au and $4.5 \km \;\lesssim \; D \; \lesssim \; 25.0 \km$. Eqs.~\ref{eq.border_method_N_outer} and \ref{eq.border_method_N_inner} are integrated using the interval ($-\infty$,$a_c\; \au$] for the Dirac delta function $\delta(a_{j}-a)$ and the interval [$0.04\;\invkm, 0.22\; \invkm$] for the Dirac delta function $\delta( D_{r,j}-D_r )$. Eq.~\ref{eqn.apvDvsC} is truncated to $0.04 \; \invkm$ for $D_r \lesssim 0.04 \; \invkm$ and to $0.22 \; \invkm$ for $D_r \gtrsim 0.22 \; \invkm$. The peak in $\frac{N_{in}}{N_{out}}$ at $(a_c, C) \; = \; (2.49 \; \mathrm{au}, \;6.6 \times 10^{-5} \; \mathrm{au})$ and corresponding V-shape are displayed in the top and bottom panels of Fig.~\ref{fig.SynthhalfB}.  The peak has a value in $\frac{N_{in}}{N_{out}}$ of more than 8 standard deviations in $\frac{N_{in}}{N_{out}}$ above the mean for the ranges 2.39 au $< \; a \;<$ 2.50 au and 1.8 $\times \; 10^{-5}$ au $< \; C \;<$ 1.0 $\times \; 10^{-4}$ au. Calculating the age of the family after revising the $6.6 \times 10^{-5} \; \mathrm{au}$ value of $C$ for the initial speed of the fragments gives an age of $\sim$800 Myr, matching the duration of the simulation. Similar results were found using the density method to locate this half V-shape family with $\sim10\%$ lower value for $C$ and family age.

\subsubsection{Two neighboring families}
\label{s.TwoV}

The ability of the V-shape method to distinguish two overlapping families was tested with a second synthetic family that was generated near the synthetic family from Section~\ref{s.SingleV}.  The second synthetic family modeled was generated at $\left (a ,e , \sin i \right ) \; = \; \left (2.28, 0.21, 0.08 \right )$ using 50,000 particles generated from a SFD with a slope of 2.85 for asteroids with $6 \km \;\lesssim \; D \; \lesssim \; 65 \km$ (where the second largest fragment was $\sim$65 km, see bottom panel of Fig~\ref{fig.SynthDoubleB}, zoomed to $0.04\;\invkm \;\lesssim \; D_r \; \lesssim \; 0.30\; \invkm$). The parent body of the second family has a diameter of 280 km, an escape speed of $\sim$ 100 $\mps$, and an initial $C$ of  $\sim 2.0 \times 10^{-5}$ au. The particles were given an additional $10$ $\mps$ ejection speed and uniformly distributed with respect to the radial, transverse and normal velocity components as for the synthetic families in Sections~\ref{s.SingleV} and \ref{s.HalfV} The semi-major axes of the second family's fragments were evolved for 800 Myr with $\sim$16,000 particles remaining at the end of the simulation. The majority of particles were removed due to collisional evolution or observational selection effects. Both families were placed together in a background of 6,000 particles uniformly distributed with a uniform distribution in $a$ and with $4.5 \km \;\lesssim \; D \; \lesssim \; 50.0 \km$ using a SFD with a slope of 2.85.

The border method was applied to both synthetic families using asteroids with 2.0 au $< \; a \;<$ 2.7 au and $4.5 \km \;\lesssim \; D \; \lesssim \; 25.0 \km$. Eqs.~\ref{eq.border_method_N_outer} and \ref{eq.border_method_N_inner} are integrated using the interval [$a_c$,$\infty$) for the Dirac delta function $\delta(a_{j}-a)$ and the interval [$0.04\;\invkm, 0.22\; \invkm$] for the Dirac delta function $\delta( D_{r,j}-D_r )$. Eq.~\ref{eqn.apvDvsC} is truncated to $0.04 \; \invkm$ for $D_r \lesssim 0.04 \; \invkm$ and to $0.22 \; \invkm$ for $D_r \gtrsim 0.22 \; \invkm$. The peak in $\frac{N_{in}}{N_{out}}$ at $(a_c, C) \; = \; (2.28 \; \mathrm{au}, \;7.5 \times 10^{-5} \; \mathrm{au})$ (Fig.~\ref{fig.SynthDoubleB}, top panel). The peak corresponding to the synthetic family in Section~\ref{s.SingleV} is visible at $(a_c, C) \; = \; (2.37 \; \mathrm{au}, \;6.5 \times 10^{-5} \; \mathrm{au})$ (Fig.~\ref{fig.SynthDoubleB}, top panel). Both peaks are more than $\sim$ 8 standard deviations above the mean value of $\frac{N_{in}}{N_{out}}$ in the range 2.0 au $< \; a \;<$ 2.7 au and 1.8 $\times \; 10^{-5}$ au $< \; C \;<$ 1.0 $\times \; 10^{-4}$ au. The results are similar when applying the density method to within $\sim 10\%$.

\subsubsection{Uniform and real main belt background}

The possibility of finding false positive V-shapes with the border and density methods was tested on 100,000 asteroids with randomly distributed using a uniform distribution of semi-major axes between 2.18 au and 2.46 au and with diameters between 5 km and 50 km using a SFD with a slope of 2.85 (see bottom panel of Fig~\ref{fig.SynthUniformD}  zoomed to $0.04\;\invkm \;\lesssim \; D_r \; \lesssim \; 0.30\; \invkm$). The top panel of Fig.~\ref{fig.SynthUniformD} shows the ratio of Eqs.~\ref{eq.border_method_N_outer} and \ref{eq.border_method_N_inner} using the interval ($-\infty$,$\infty$) for the Dirac delta function $\delta(a_{j}-a)$ and the interval [$0.04\;\invkm, 0.22\; \invkm$] for the Dirac delta function $\delta( D_{r,j}-D_r )$. A smooth distribution with no $\frac{N_{in}}{N_{out}}$ values significantly greater than 1 in $a_c$ vs. $C$ space suggesting that the border method is does not find false V-shapes in uniformly randomized data. The density method gives a similar result using the same intervals. Intervals for half V-shapes, ($-\infty$,$a_c$] and [$a_c$,$\infty$), were also used with the border and density methods are also applied to the uniform background and give a similar smooth distribution in $a_c$ vs. $C$ space as the top panel of Fig.~\ref{fig.SynthUniformD}.

The border method was tested on a section of the main belt with no family V-shapes.  1823 asteroids were used with $3.00 \au \; < \; a \; < \; 3.25 \au$, $0.00 \; < \; e \; < \; 0.12$, $0.00 \; < \; \sin i \; < \; 0.12$ and $0.01 \; < \; \pv \; < \; 0.30$. The top panel of Fig.~\ref{fig.RealMBD} shows the ratio of Eqs.~\ref{eq.border_method_N_outer} and \ref{eq.border_method_N_inner} using the interval ($-\infty$,$\infty$) for the Dirac delta function $\delta(a_{j}-a)$ and the interval [$0.04\;\invkm, 0.22\; \invkm$] for the Dirac delta function $\delta( D_{r,j}-D_r )$. A smooth distribution with the majority of $\frac{N_{in}}{N_{out}}$ values approximately equal to 1. The peak value of $\frac{N_{in}}{N_{out}}$ is $\sim$2 near $a_c \; = \; 3.23$ and $C \; = \; 2.0 \times 10^{-5}$ au does not correspond to any known family V-shape.

\subsection{Test of the methods on real families}

The inner belt is sculpted by several powerful resonances, which are $a$,  $e$ and $i$ dependent that affect asteroid families as described in Section~\ref{s.Introduction}. Examples include the inclination-dependent $\nu_6$ resonance at the inner boundary of the main belt sculpts the Flora family \citep[see Fig. 15 of ][]{Milani1990} and the 3:1 MMR with Jupiter sculpts the Eulalia and new Polana  families at its inner edge in $a$ and at increased eccentricities \citep[see Figs. 18 and 19 of ][]{Wisdom1983}. Collisional families affected by these inner main belt resonances are ideal for testing the robustness of V-shape finding techniques since their V-shape differs from the standard V-shape. The V-shape finding method is first tested out on families more easily identified with the V-shape method moving to more families more difficult to identify with the V-shape technique. 

\subsection{Data set}

The data set used to test the V-shape technique on real asteroid families includes diameter measurements from the  WISE catalogue \cite[][]{Wright2010,Mainzer2011b,Masiero2011} for 102,400 MBAs. Only diameter measurements, which have $<\;30\%$ relative uncertainty from the WISE catalogue were included in the data set. Absolute magnitude measurements of 66,655 MBAs from the PanSTARRS photometric catalogue \citep[][]{Kaiser2010,Denneau2013,Veres2015} that had a photometric uncertainty of less than 0.1 magnitudes were used that also had diameter measurements from the WISE catalogue. The average relative uncertainty of absolute magnitudes from the PanSTARRS catalogue is $\sim$0.04 magnitudes \citep[][]{Veres2015}. Absolute magnitude measurements were taken from the MPC catalogue, which did not have an absolute magnitude measurements from the PanSTARRS catalogue. Synthetic MBA proper elements were taken from Asteroid Dynamic Site\footnote{\tt{http://hamilton.dm.unipi.it/astdys/}} \citep[see Fig.~\ref{fig.innerbelt},][]{Knezevic2003}. Numerical proper elements were used preferentially and analytical proper elements were used for asteroids, which did have numerically calculated elements as of April 2016.

\subsubsection{Erigone}

The young Erigone family is an example of a collisional family with a complete V-shape (see the bottom panel of Fig.~\ref{fig.EulaliaD}) and has an age between 200 and 300 Myr \citep[][]{Vokrouhlicky2006b,Broz2013b, Spoto2015,Bottke2015a}. 

The V-shape density technique is enhanced when identifying the Erigone family by using $H$ magnitudes from the PanSTARRS and MPC catalogues compared to when using diameter measurements from the WISE catalogue because the H magnitudes in the sample from the PanSTARRS catalogue are more accurate than the diameter measurements from the WISE catalogue for the Erigone family. The density technique is applied to 715 asteroids with proper elements $2.26 \au \; < \; a \; < \; 2.47 \au$, $0.20 \; < \; e \; < \; 0.22$, $0.08 \; < \; \sin i \; < \; 0.11$ and $0.01 \; < \; \pv \; < \; 0.10$ as defined for the Erigone family by \citet{Masiero2013}. Asteroids with $\pv$ between 0.01 and 0.1 are used because the majority of asteroids in the Erigone family are C-type asteroids \citep[][]{Spoto2015,Nesvorny2015a}. Asteroids with $H$ magnitudes between 12.8 (the brightest asteroid in the proper elements and $\pv$ ranges described above) and 17 are used.

\begin{figure}
\centering
\hspace*{-2.7cm}
\ifincludeplots
\ifgrayscale
\includegraphics[scale=0.50]{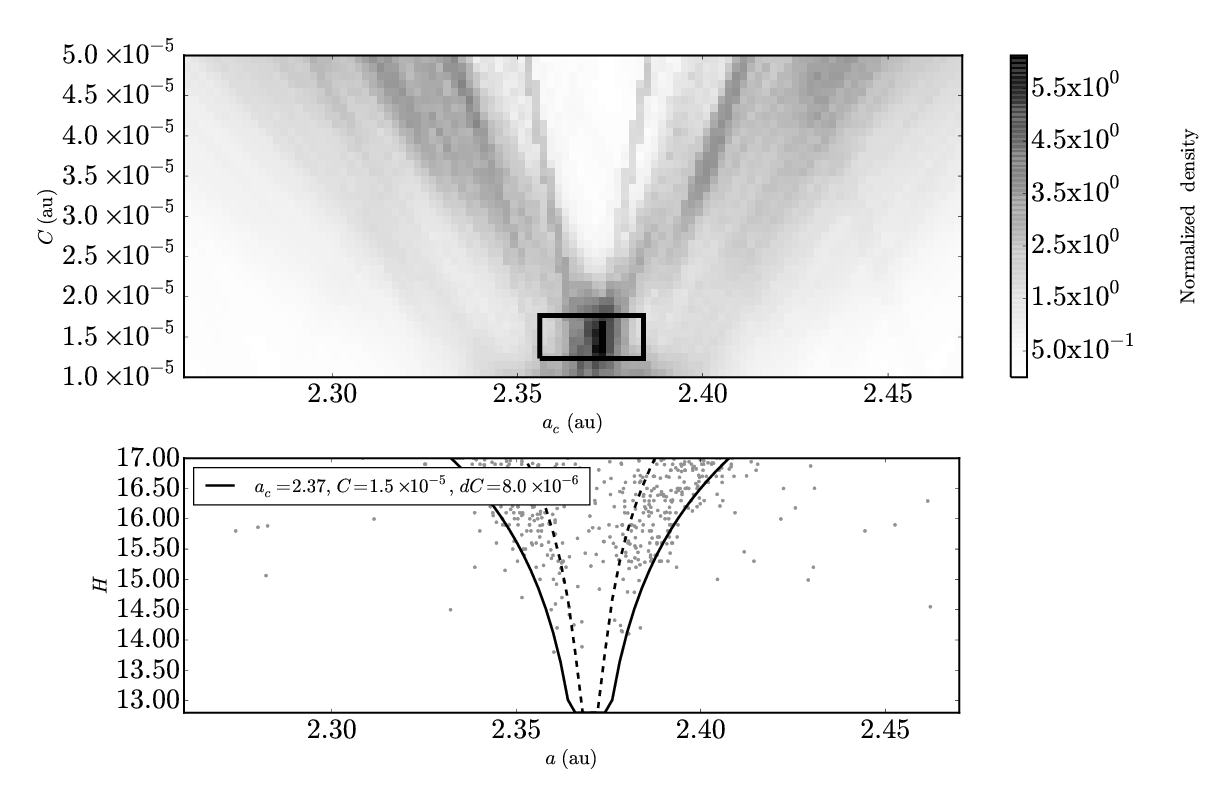}
\else
\includegraphics[scale=0.50]{Erigone_density.png}
\fi
\else
I am not enabling plots.
\fi
\caption{The density method applied to the Erigone family V-shape. (Top panel) The normalized density in units of au$^{-1}$ for asteroids in the inner V-shape in the $a_c$-$C$ range, ($a_c\pm \frac{\Delta a_c}{2}$,$C\pm \frac{\Delta C}{2}$) where $\Delta a_c \; = \; 2.0 \times 10^{-3}$ au and $\Delta C \; = \; 1.0 \times 10^{-6}$ au for the Erigone family. The box marks the peak value in the normalized density for the synthetic family V-shape. (Bottom Panel) $H(a,a_c,C)$ is plotted for the peak values with the primary V-shape as a solid line. The dashed line mark the boundary for the area in $a$ vs. $H$ space for the inner V-shape using Eq.~\ref{eqn.aHvsC},  $H(a,a_c,C- dC)$ where $dC \; = \; 8.0 \x 10^{-6}$ au. The picket fence pattern in $H$ axis direction is an artifact cause by the inclusion of MPC $H$ magnitudes in which the majority have a precision of 0.1 magnitudes.}
\label{fig.ErigoneD}
\end{figure}

Eq.~\ref{eq.density_method} is integrated using the interval ($-\infty$,$\infty$) for the Dirac delta function $\delta(a_{j}-a)$ and the interval [$12.8,17.0$] for the Dirac delta function $\delta(H_j-H )$. Eq.~\ref{eqn.aHvsC} is truncated to 12.8 for $H$ $<$ 12.8 and to 17 for $H$ $ >$ $17.0$. The peak in $\rho$ at $(a_c, C) \; = \; (2.37 \; \mathrm{au}, \;1.5 \times 10^{-5} \; \mathrm{au})$ (Fig.~\ref{fig.ErigoneD}, top panel) corresponds the to location of the family V-shape (bottom panel of Fig.~\ref{fig.ErigoneD}). The peak value of $\rho$ is  $\sim$5 standard deviations above the mean value of $\rho$ in the range 2.26 au $< \; a \;<$ 2.47 au and 1.0 $\times \; 10^{-5}$ au $< \; C \;<$ 5.0 $\times \; 10^{-5}$ au.  A $dC \; = \; 8.0 \; \times 10^{-6}$ au was used. The value of $C \; = \; 1.5 \; \times 10^{-5} \; \au$ is in good agreement with the value reported by \citet[][]{Nesvorny2015a} (see Table 2) suggesting that the V-shape found with the density method is a good match.

\subsubsection{Flora and Baptistina}

The Flora family was used as a test for the robustness of the V-shape to diffusion in $e$ and $i$ on Gyr caused by numerous resonances on Gyr-time scales \citep[][]{Milani1994b,Nesvorny2002a} given its age of $\sim$ 950 Myr \citep[][]{Dykhuis2014}. There is non-agreement in Flora's definition as a collisional family because it is not found with the HCM techniques of \citet{Milani2014}, but is found in other recent work by different versions of HCM \citep[][]{Dykhuis2014,Nesvorny2015a}.

Exactly 2399 Asteroids with proper elements $2.16 \au \; < \; a \; < \; 2.40 \au$, $0.10 \; < \; e \; < \; 0.18$, $0.05 \; < \; \sin i \; < \; 0.13$ and $0.20 \; < \; \pv \; < \; 0.38$ as defined for the Flora family by \citet{Dykhuis2014} are used. The density technique is enhanced when identifying the Flora family by using H magnitude measurements from the PanSTARRS and MPC compared to diameter measurements from the WISE catalogue because the H magnitudes in the sample from the PanSTARRS catalogue are more accurate than the diameter measurements from the WISE catalogue for the Flora family. The inner side of the Flora family is heavily sculpted by the $\nu_6$ resonance \citep[][]{Nesvorny2002a}. The outer V-shape of the Flora family is not affected by the resonance, so Eq.~\ref{eq.density_method} is integrated using the interval [$a_c$,$\infty$) for the Dirac delta function $\delta(a_{j}-a)$. The interval [$10.0, 18.0$] was used for the Dirac delta function $\delta(H_j-H )$. Eq.~\ref{eqn.aHvsC} is truncated to 10.0 for $H$ $ <$ 10.0 and to 18 for $H$ $>$ 18.0. The lower bound of 11.0 in $H$ was used because an $H$ of 11.0 corresponds to a a diameter of $\sim$ 16 km assuming a $\pv$ of 0.29 and is equivalent to the lower limit of $H \; >\;12.0$ assuming $\pv$ as described in Section~\ref{s.Vshapeidentification}. The peak in $\rho$ is located at $(a_c, C) \; = \; (2.205 \; \mathrm{au}, \;1.5 \times 10^{-4} \; \mathrm{au})$ (Fig.~\ref{fig.FloraD}, top panel) and a $dC \; = \; 3.2 \; \times 10^{-5}$ au was used. The value of $C \; = \; 1.5 \; \times 10^{-4} \; \au$ is similar to the value of $C \; = \; 1.7 \; \times 10^{-4} \; \au$ found in \citet{Dykhuis2014}. The peak value of $\rho$ is $\sim$ 4 standard deviations above the mean in the range 2.16 au $< \; a \;<$ 2.70 au and 3.3 $\times \; 10^{-5}$ au $< \; C \;<$ 4.0 $\times \; 10^{-4}$ au.

The $\sim$ 160 Myr-old Baptistina family is recognized by its V-shape in $a$ vs. $H$ space within the HCM-defined Flora family \citep[][]{Bottke2007,Nesvorny2015a}. 3912 asteroids were used with $2.16 \au \; < \; a \; < \; 2.40 \au$, $0.10 \; < \; e \; < \; 0.18$, $0.05 \; < \; \sin i \; < \; 0.13$, identical to the orbital elements used for the Flora family,  and $0.1 \; < \; \pv \; < \; 0.38$ since this range in $\pv$ will include both the Flora and Baptistina family \citep[][]{Reddy2009,Spoto2015}. 

The Baptistina family is identified in $a$ vs. $D_r$ space with the density method (Fig.~\ref{fig.BaptistinaD}). Eq.~\ref{eq.density_method} is integrated using the interval [$a_c$,$\infty$) for the Dirac delta function $\delta(a_{j}-a)$ because the V-shape of the Baptistina family is bisected by the the 7:2 MMR with Jupiter leaving the outer V-shape half mostly intact. The interval [$0.19\;\invkm, 1.00\; \invkm$] was used for the Dirac delta function $\delta(D_{r,j}-D_r )$. Eq.~\ref{eqn.apvDvsC} is truncated to $0.19 \; \invkm$ for $D_r \lesssim 0.19 \; \invkm$ and to $1.0 \; \invkm$ for $D_r \gtrsim 1.0 \; \invkm$. The peak in $\rho$ is located at $(a_c, C) \; = \; (2.265 \; \mathrm{au}, \;2.4 \times 10^{-4} \; \mathrm{au})$ (Fig.~\ref{fig.BaptistinaD}, top panel). A smaller value for $dC$ was used, $dC \; = \; 8.0 \; \times 10^{-6}$ au, compared to the value of $dC$ used for the Flora family V-shape because the Baptistina family V-shape edges are more dense than the V-shape edges for the Flora family. The Baptistina family is also younger than the Flora family . The value of $C \; = \; 2.4 \; \times 10^{-5} \; \au$ is similar to the values of $C \; = \; 1.5 \; \times 10^{-5} \; \au$ and  $C \; = \; 2.5 \; \times 10^{-5} \; \au$ found for the Baptistina family in \citet[][]{Bottke2007} and \citet[][]{Nesvorny2015a}. The peak value of $\rho$ is $\sim$ 4 standard deviations above the mean in the range 2.16 au $< \; a \;<$ 2.40 au and 1.0 $\times \; 10^{-5}$ au $< \; C \;<$ 7.0 $\times \; 10^{-5}$ au. 

\subsubsection{Vesta}

The Vesta family may be the result of two cratering events \citep[][]{Farinella1996,Milani2014} corresponding to the creation of the $\sim$1 Gyr-old Rheasilvia basin \citep[][]{Marchi2012}  and the $\sim$2 Gyr-old Veneneia basin \citep[][]{Obrien2014}.  The border method is enhanced when identifying the Vesta family using diameter measurements from the WISE catalogue is compared to $H$ measurements from the PanSTARRS and MPC catalogues. Exactly 1902 asteroids with proper elements $2.25 \au \; < \; a \; < \; 2.5 \au$, $0.07 \; < \; e \; < \; 0.14$, $0.09 \; < \; \sin i \; < \; 0.14$ and $0.15 \; < \; \pv \; < \; 0.60$ as defined for the family by \citet{Milani2014} and \citet{Spoto2015} are used. Eqs.~\ref{eq.border_method_N_outer} and \ref{eq.border_method_N_inner} are integrated using the interval ($-\infty$,$\infty$) for the Dirac delta function $\delta(a_{j}-a)$ and the interval [$0.2\;\invkm, 0.70\; \invkm$] for the Dirac delta function $\delta( D_{r,j}-D_r )$. A $dC \; = \; 3.2 \; \times10^{-5}$ au was used. Eq.~\ref{eqn.apvDvsC} is truncated to $0.2 \; \invkm$ for $D_r \lesssim 0.2 \; \invkm$ and to $0.70 \; \invkm$ for $D_r \gtrsim 0.70 \; \invkm$. A higher weight of 4.0 was used in Eqs.~\ref{eq.border_method_N_outer} and \ref{eq.border_method_N_inner} corresponding to a higher SFD slope expected of family fragments produced by cratering events \citep[][]{Tanga1999,Bottke2005a}. The peak in $\frac{N_{in}}{N_{out}}$ at $(a_c, C) \; = \; (2.37 \; \mathrm{au}, \;1.4 \times 10^{-4} \; \mathrm{au})$ (Fig.~\ref{fig.VestaB}, top panel) is similar to the value of $C \; = \; 1.5 \; \times 10^{-4} \; \au$ found in \citet{Nesvorny2015a}. The peak value of  $\frac{N_{in}}{N_{out}}$  found with the border method is $\sim$18 standard deviations above the mean value of  $\frac{N_{in}}{N_{out}}$  in the range 2.25 au $< \; a \;<$ 2.5 au and 3.2 $\times \; 10^{-5}$ au $< \; C \;<$ 4.0 $\times \; 10^{-4}$ au. A statistically significant peak corresponding to a possible second, older family was not found.

\subsubsection{New Polana  and Eulalia}

New Polana  and Eulalia are incomplete or half V-shape families located near the 3:1 MMR with Jupiter and are inseparable with HCM and have similar C and B-type spectra \citep[][]{Walsh2013, Dykhuis2015, Pinilla-Alonso2016}. New Polana and Eulalia were once identified as a single family named after Polana, the latter being part of a larger cluster dubbed the Nysa-Polana cluster \citep[][]{Cellino2002a,Mothe-Diniz2005,Campins2010}. The separation and definition of the new Polana  and Eulalia families was made by identifying their half V-shape \citep[][]{Walsh2013}. 

\begin{figure}
\centering
\hspace*{-2.7cm}
\ifincludeplots
\ifgrayscale
\includegraphics[scale=0.50]{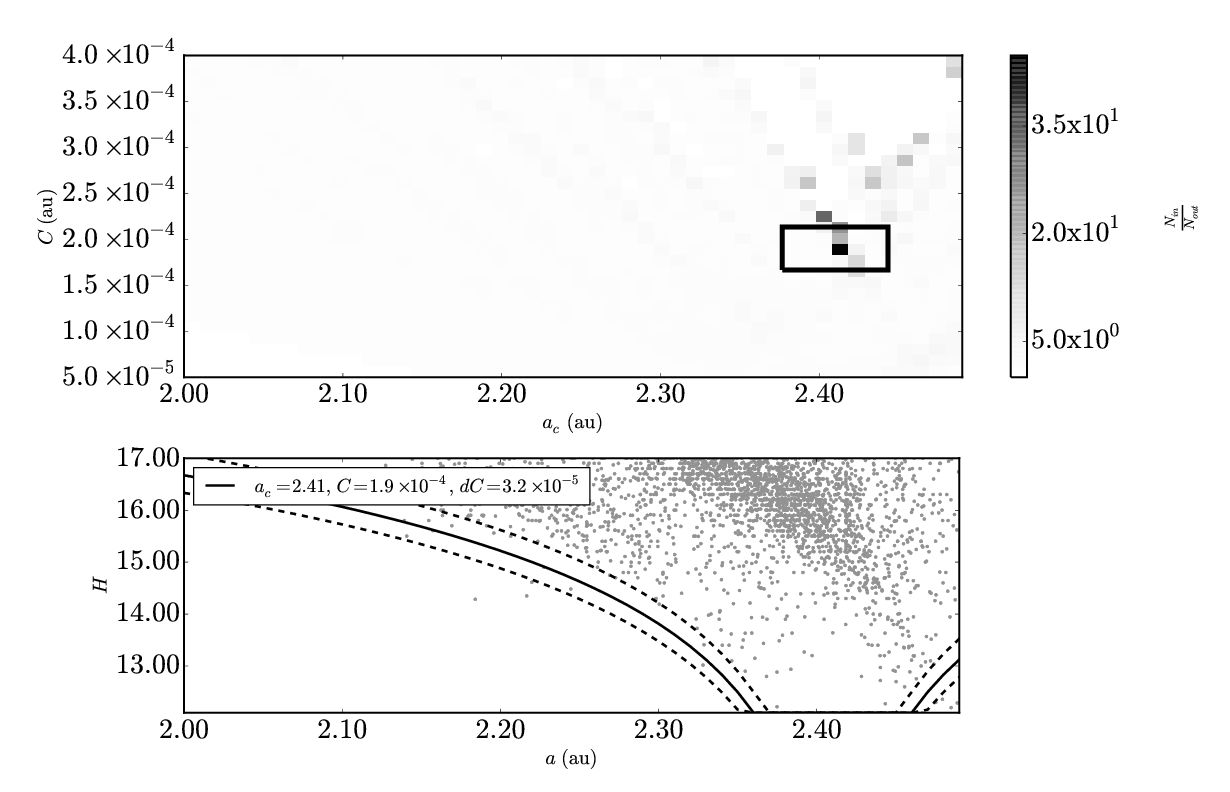}
\else
\includegraphics[scale=0.50]{Polana_border.png}
\fi
\else
I am not enabling plots.
\fi
\caption{The border method applied to the Polana family V-shape. (Top panel) The ratio between the number of asteroids in the outer V-shape to the number of asteroids in the inner V-shape in the $a_c$-$C$ range, ($a_c\pm \frac{\Delta a_c}{2}$,$C\pm \frac{\Delta C}{2}$) where $\Delta a_c \; = \; 5.0 \times 10^{-3}$ au and $\Delta C \; = \; 6.0 \times 10^{-6}$ au for the new Polana  family. The box marks the peak value in $\frac{N_{out}(a_c,C,dC)}{N_{in}(a_c,C,dC)}$ for the synthetic family V-shape. (Bottom Panel) $D_r(a,a_c,C,\pv)$ is plotted for the peak values with the primary V-shape as a solid line where $\pv = 0.05$. The dashed lines mark the boundaries for the area in $a$ vs. $D_r$ space for $N_{in}$ and $N_{out}$ using Eq.~\ref{eqn.apvDvsC},  $D_r(a,a_c,C\pm dC,\pv)$ where $dC \; = \; 3.2 \x 10^{-5}$ au.}
\label{fig.NewPolanaB}
\end{figure}

Exactly 3578 asteroids with with proper elements $2.0 \au \; < \; a \; < \; 2.5 \au$, $0.1 \; < \; e \; < \; 0.2$, $0.02 \; < \; \sin i \; < \; 0.09$ and $0.01 \; < \; \pv \; < \; 0.10$ as defined for the new Polana  by \citet{Walsh2013} are used. The border technique is enhanced when identifying the new Polana  family by using H magnitude measurements from the PanSTARRS and MPC compared to diameter measurements from the WISE catalogue because the H magnitudes in the sample from the PanSTARRS catalogue are more accurate than the diameter measurements from the WISE catalogue for the new Polana  family.  Eqs.~\ref{eq.border_method_N_outer} and \ref{eq.border_method_N_inner} are integrated using the interval ($-\infty$,$\infty$) for the Dirac delta function $\delta(a_{j}-a)$ and the interval [$0.05\;\invkm, 0.70\; \invkm$] for the Dirac delta function $\delta( D_{r,j}-D_r )$. A $dC \; = \; 3.2 \; \times10^{-5}$ au was used. Eq.~\ref{eqn.apvDvsC} is truncated to $12.0$ for $H \; <  \;12.0$ and to $17.0$ for $H \; > \; 17.0$. The peak in $\frac{N_{in}}{N_{out}}$ at $(a_c, C) \; = \; (2.4 \; \mathrm{au}, \;2.0 \times 10^{-4} \; \mathrm{au})$ (Fig.~\ref{fig.NewPolanaB}, top panel) is similar to the value of $C \; = \; 1.7 \; \times 10^{-4} \; \au$ found for new Polana  in \citet{Walsh2013}. The peak value of  $\frac{N_{in}}{N_{out}}$  found with the border method is $\sim$12 standard deviations above the mean value of  $\frac{N_{in}}{N_{out}}$  in the range 2.0 au $< \; a \;<$ 2.5 au and 5.0 $\times \; 10^{-5}$ au $< \; C \;<$ 4.0 $\times \; 10^{-4}$ au. The corresponding V-shape found for the new Polana  family is plotted in the bottom panel of Fig.~\ref{fig.NewPolanaB} zoomed to $12.0 \; < \; H \; < \; 17.0$.

\begin{figure}
\centering
\hspace*{-2.7cm}
\ifincludeplots
\ifgrayscale
\includegraphics[scale=0.50]{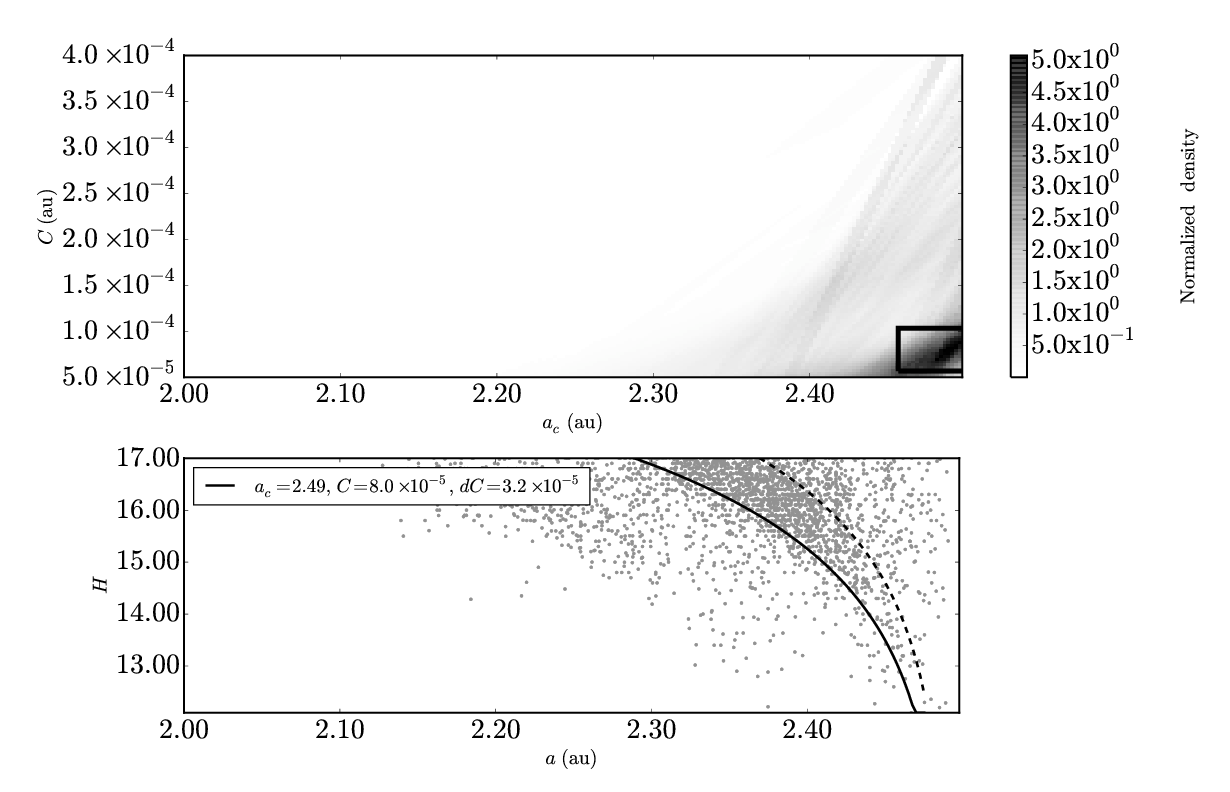}
\else
\includegraphics[scale=0.50]{Eulalia_density.png}
\fi
\else
I am not enabling plots.
\fi
\caption{The density method applied to the Eulalia family V-shape.Ê(Top panel) The normalized density in units of au$^{-1}$ for asteroids in the inner V-shape in the $a_c$-$C$ range, ($a_c\pm \frac{\Delta a_c}{2}$,$C\pm \frac{\Delta C}{2}$) where $\Delta a_c \; = \; 2.5 \times 10^{-3}$ au and $\Delta C \; = \; 3.0 \times 10^{-6}$ au for the Eulalia family. The box marks the peak value in the normalized density for the synthetic family V-shape. (Bottom Panel) $H(a,a_c,C)$ is plotted for the peak values with the primary V-shape as a solid line. The dashed line mark the boundary for the area in $a$ vs. $H$ space for the inner V-shape using Eq.~\ref{eqn.aHvsC},  $H(a,a_c,C- dC)$ where $dC \; = \; 3.2 \x 10^{-5}$ au. The picket fence pattern in $H$ axis direction is an artifact caused by the inclusion of MPC $H$ magnitudes of, which the majority have a precision of 0.1 magnitudes.}
\label{fig.EulaliaD}
\end{figure}

\begin{figure}
\centering
\hspace*{-2.7cm}
\ifincludeplots
\ifgrayscale
\includegraphics[scale=0.50]{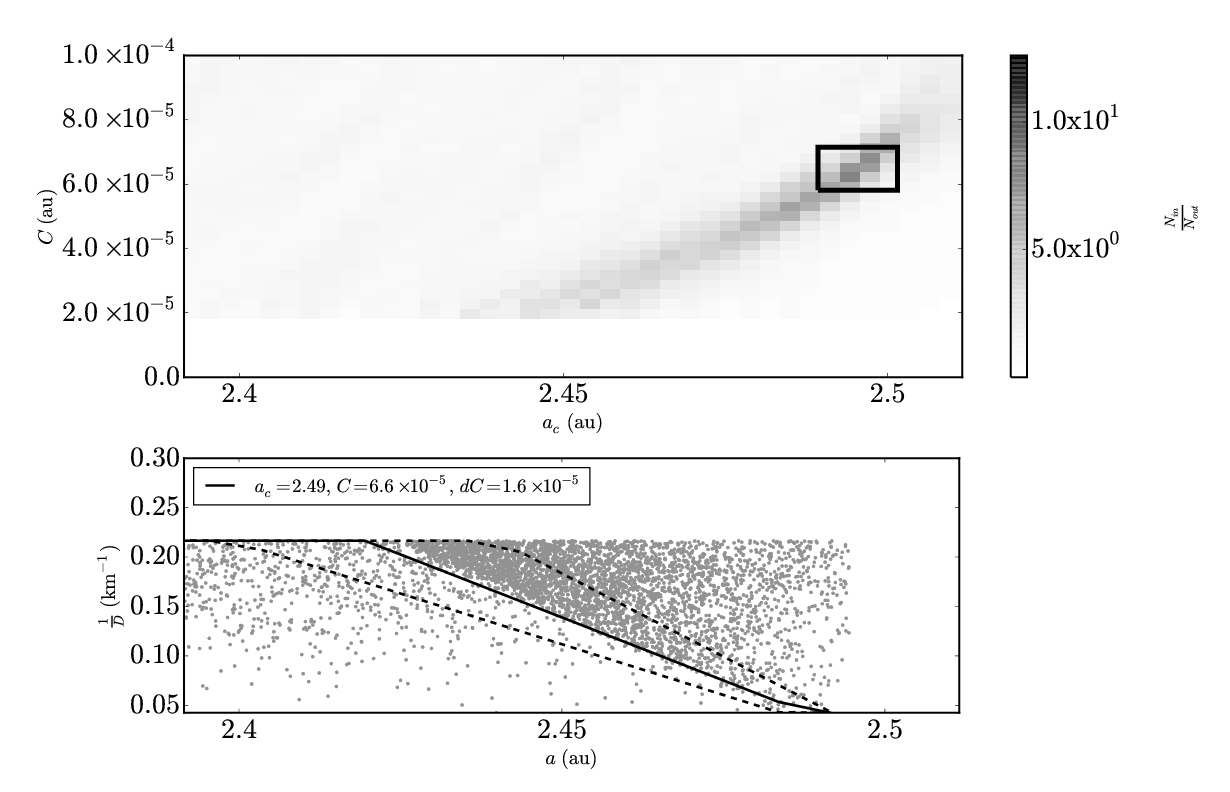}
\else
\includegraphics[scale=0.50]{Synth_half_v_shape_border.png}
\fi
\else
I am not enabling plots.
\fi
\caption{Application of the border method on a synthetic half V-shape family. (Top panel) The ratio between the number of asteroids in the outer V-shape to the number of asteroids in the inner V-shape in the $a_c$-$C$ range, ($a_c\pm \frac{\Delta a_c}{2}$,$C\pm \frac{\Delta C}{2}$) where $\Delta a_c \; = \; 10.0 \times 10^{-3}$ au and $\Delta C \; = \; 12.0 \times 10^{-6}$ au for a single synthetic family with a half-V-shape. The box marks the peak value in $\frac{N_{out}(a_c,C,dC)}{N_{in}(a_c,C,dC)}$ for the synthetic family V-shape. (Bottom Panel) $D_r(a,a_c,C,\pv)$ is plotted for the peak values with the primary V-shape as a solid line where $\pv = 0.05$. The dashed lines mark the boundaries for the area in $a$ vs. $D_r$ space for $N_{in}$ and $N_{out}$ using Eq.~\ref{eqn.apvDvsC},  $D_r(a,a_c,C\pm dC,\pv)$ where $dC \; = \; 1.6 \x 10^{-5}$ au.}
\label{fig.SynthhalfB}
\end{figure}

\begin{figure}
\centering
\hspace*{-2.7cm}
\ifincludeplots
\ifgrayscale
\includegraphics[scale=0.50]{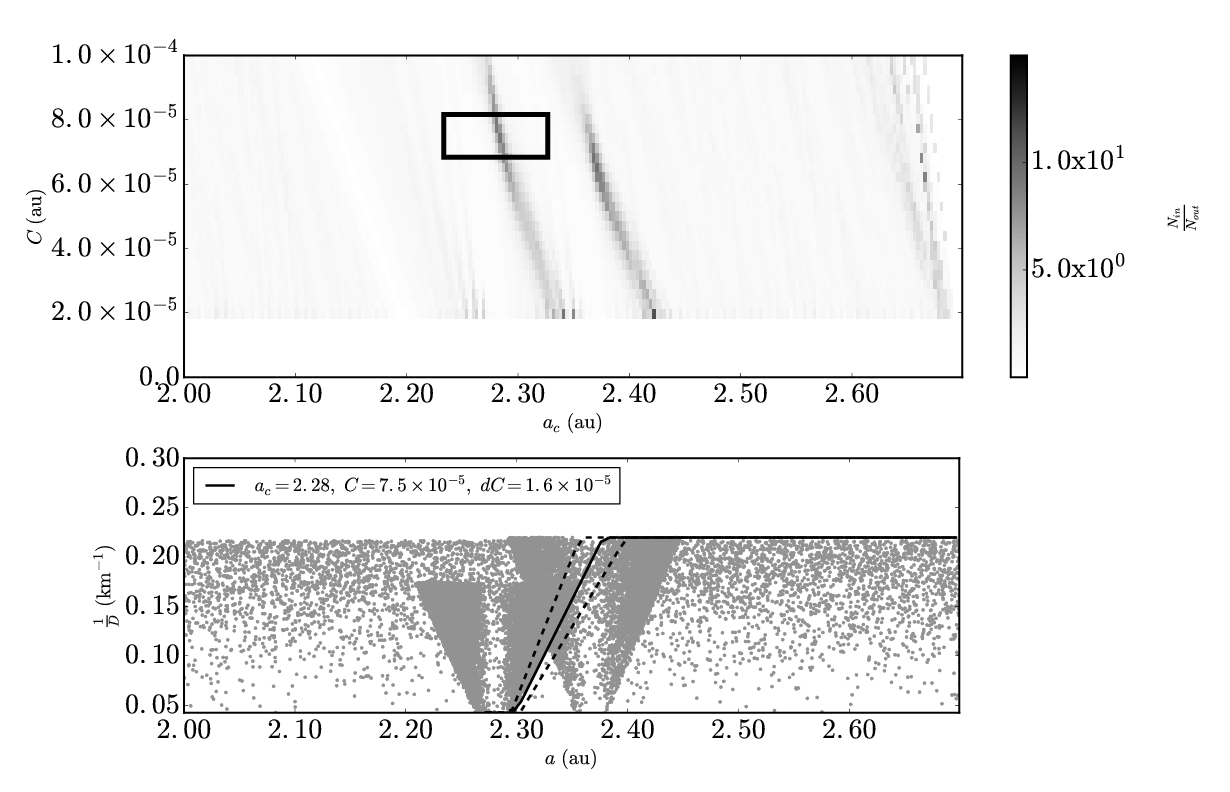}
\else
\includegraphics[scale=0.50]{Synth_double_border.png}
\fi
\else
I am not enabling plots.
\fi
\caption{Application of the border method on two adjacent family V-shapes.The same as in Fig.~\ref{fig.SynthsingleB} including an additional synthetic family at $a_c\; = \; 2.28$ au and using a half V-shape. There are no asteroids beyond 2.7 au which artificially raises $N_{in}$/$N_{out}$ when integrating Eqs.~\ref{eq.border_method_N_outer} and \ref{eq.border_method_N_inner} between [$a_c$,$\infty$) for the Dirac delta function $\delta(a_{j}-a)$  causing a small artifact near 2.65 au.}
\label{fig.SynthDoubleB}
\end{figure}

\begin{figure}
\centering
\hspace*{-2.7cm}
\ifincludeplots
\ifgrayscale
\includegraphics[scale=0.50]{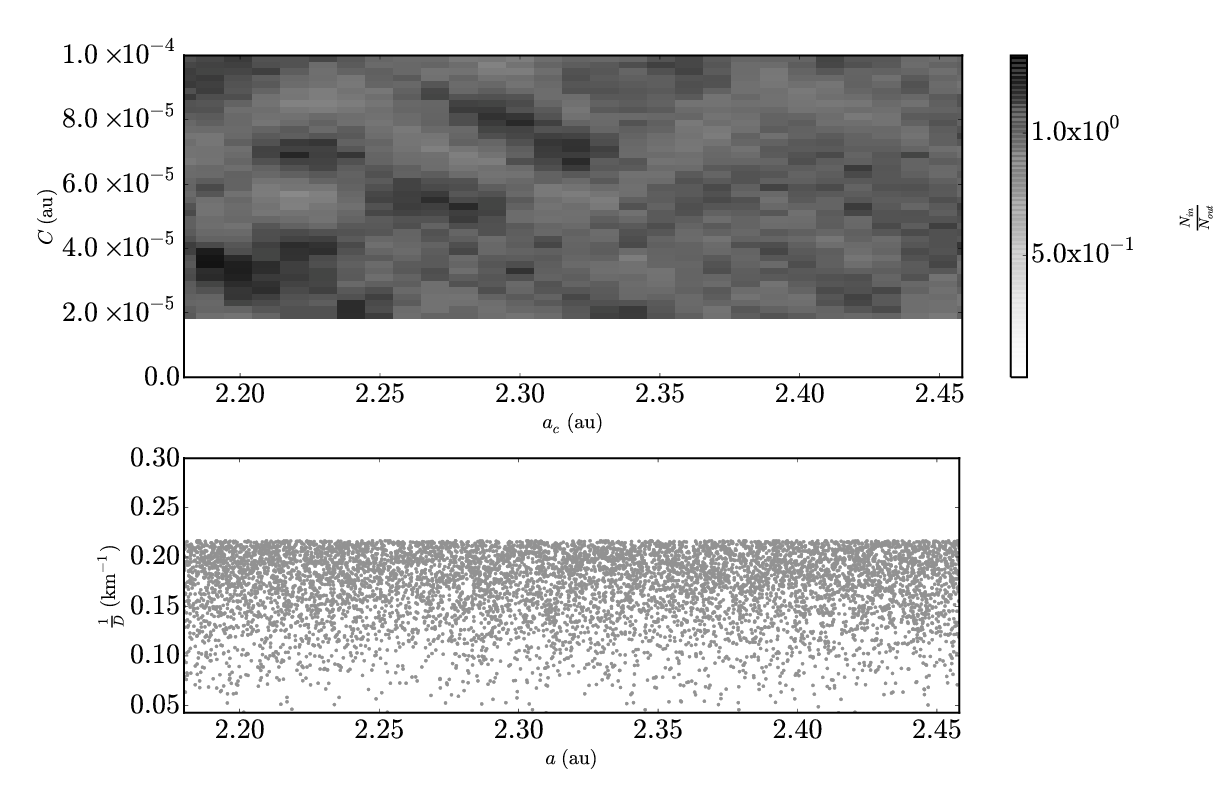}
\else
\includegraphics[scale=0.50]{Synth_uniform_border.png}
\fi
\else
I am not enabling plots.
\fi
\caption{Application of the border method on a uniformly random background of asteroids. (Top panel) The ratio between the number of asteroids in the outer V-shape to the number of asteroids in the inner V-shape in the $a_c$-$C$ range, ($a_c\pm \frac{\Delta a_c}{2}$,$C\pm \frac{\Delta C}{2}$) where $\Delta a_c \; = \; 2.0 \times 10^{-3}$ au, $\Delta C \; = \; 2.0 \times 10^{-6}$ au and $dC \; = \; 1.6 \x 10^{-5}$ au for a uniform background (bottom panel).}
\label{fig.SynthUniformD}
\end{figure}

\begin{figure}
\centering
\hspace*{-2.7cm}
\ifincludeplots
\ifgrayscale
\includegraphics[scale=0.50]{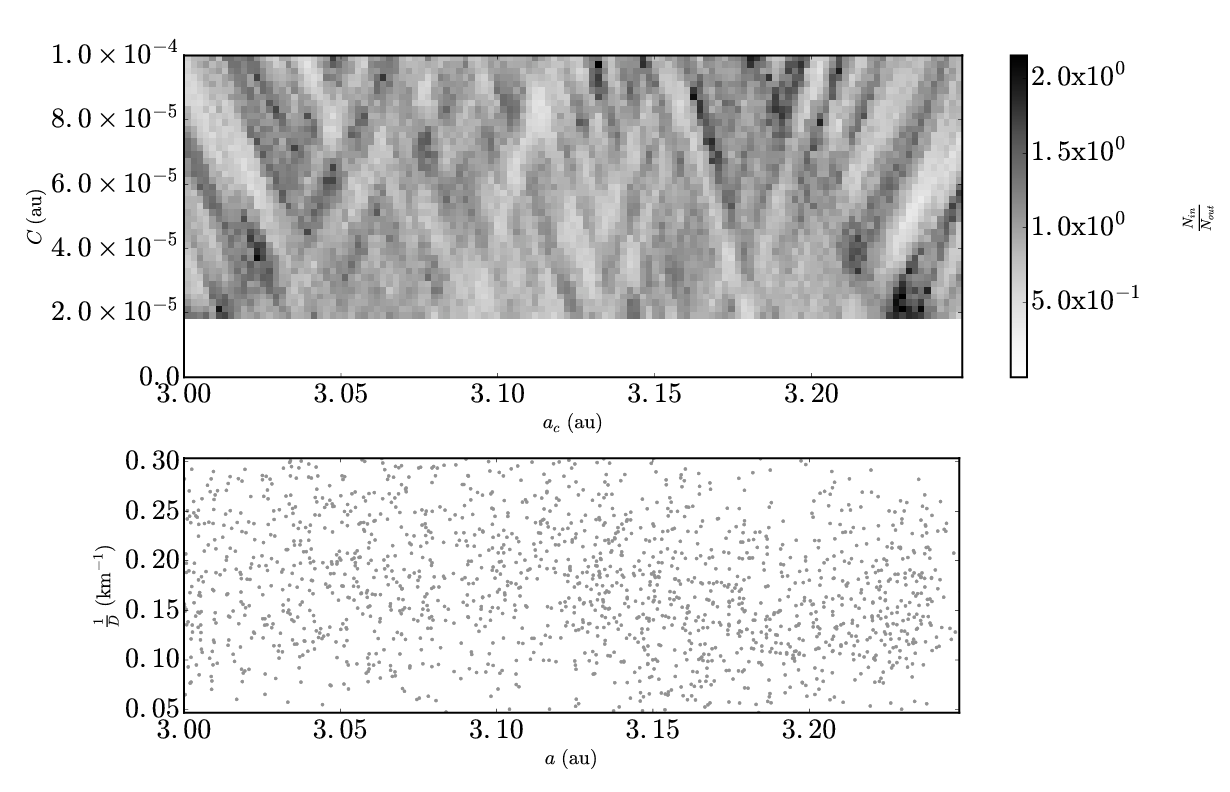}
\else
\includegraphics[scale=0.50]{Main_belt_border.png}
\fi
\else
I am not enabling plots.
\fi
\caption{Application of the border method on a section of the main belt background of asteroids. (Top panel) The ratio between the number of asteroids in the outer V-shape to the number of asteroids in the inner V-shape in the $a_c$-$C$ range, ($a_c\pm \frac{\Delta a_c}{2}$,$C\pm \frac{\Delta C}{2}$) where $\Delta a_c \; = \; 2.0 \times 10^{-3}$ au, $\Delta C \; = \; 2.0 \times 10^{-6}$ au and $dC \; = \; 1.6 \x 10^{-5}$ au for the main belt background (bottom panel).}
\label{fig.RealMBD}
\end{figure}

\begin{figure}
\centering
\hspace*{-2.7cm}
\ifincludeplots
\ifgrayscale
\includegraphics[scale=0.425]{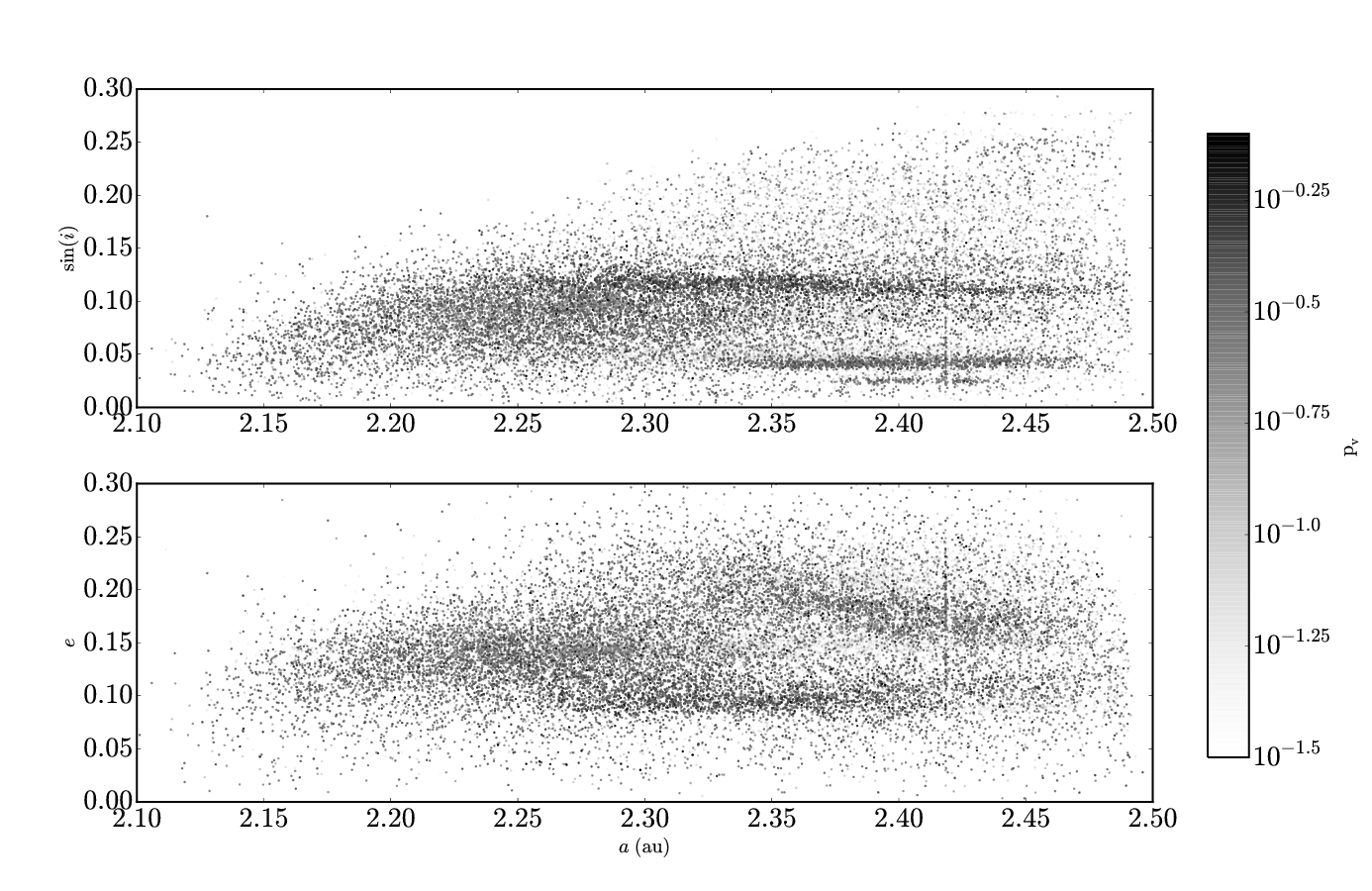}
\else
\includegraphics[scale=0.425]{inner_belt.png}
\fi
\else
I am not enabling plots.
\fi
\caption{Proper elements distribution of inner main belt asteroids. The color scale is the geometric albedo $\pv$ calculated from diameters from \citet{Masiero2011}.}
\label{fig.innerbelt}
\end{figure}

\begin{figure}
\centering
\hspace*{-2.7cm}
\ifincludeplots
\ifgrayscale
\includegraphics[scale=0.50]{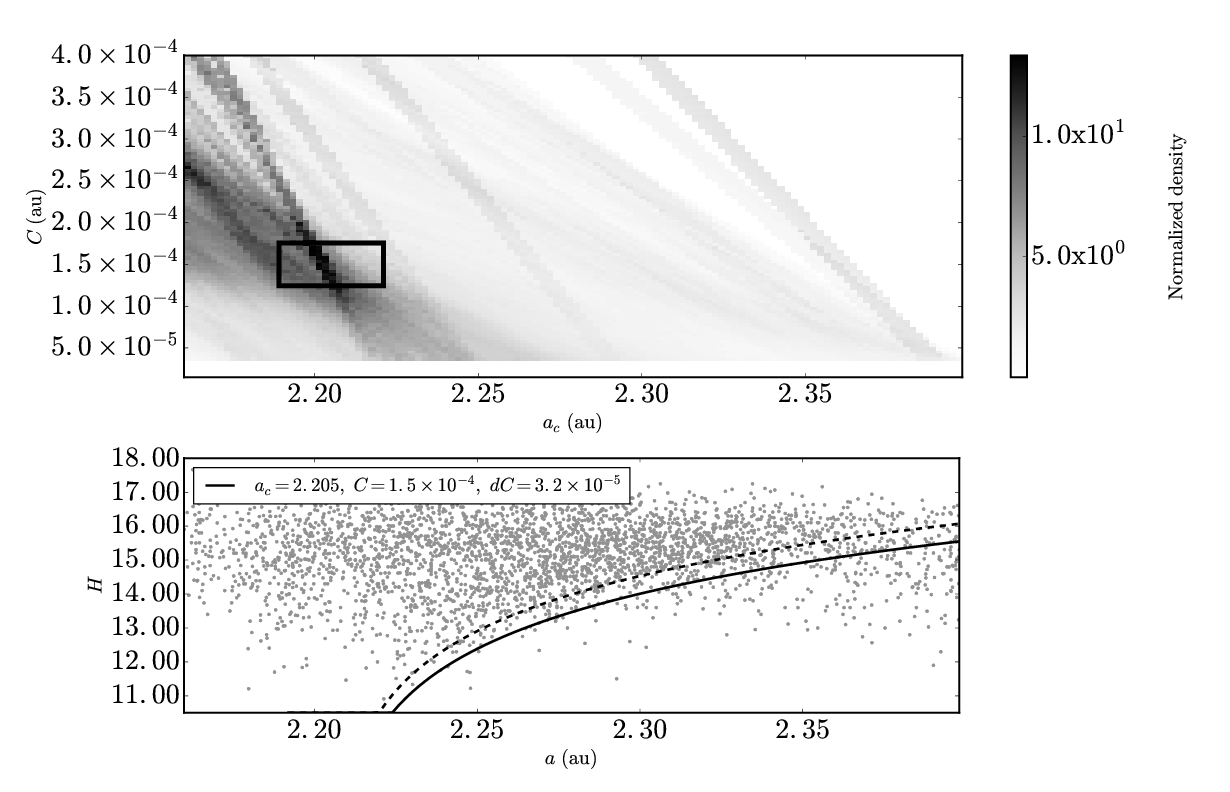}
\else
\includegraphics[scale=0.50]{Flora_density.png}
\fi
\else
I am not enabling plots.
\fi
\caption{The density method applied to the Flora family V-shape. (Top panel) The normalized density in units of au$^{-1}$ for asteroids in the inner V-shape in the $a_c$-$C$ range, ($a_c\pm \frac{\Delta a_c}{2}$,$C\pm \frac{\Delta C}{2}$) where $\Delta a_c \; = \; 2.0 \times 10^{-3}$ au and $\Delta C \; = \; 4.0 \times 10^{-6}$ au for the Flora family. The box marks the peak value in the normalized density for the synthetic family V-shape. (Bottom Panel) $H(a,a_c,C)$ is plotted for the peak values with the primary V-shape as a solid line. The dashed line mark the boundary for the area in $a$ vs. $H$ space for the inner V-shape using Eq.~\ref{eqn.aHvsC},  $H(a,a_c,C- dC)$ where $dC \; = \; 3.2 \x 10^{-5}$ au. The picket fence pattern in $H$ axis direction is an artifact caused by the inclusion of MPC $H$ magnitudes of, which the majority have a precision of 0.1 magnitudes.}
\label{fig.FloraD}
\end{figure}

\begin{figure}
\centering
\hspace*{-2.7cm}
\ifincludeplots
\ifgrayscale
\includegraphics[scale=0.50]{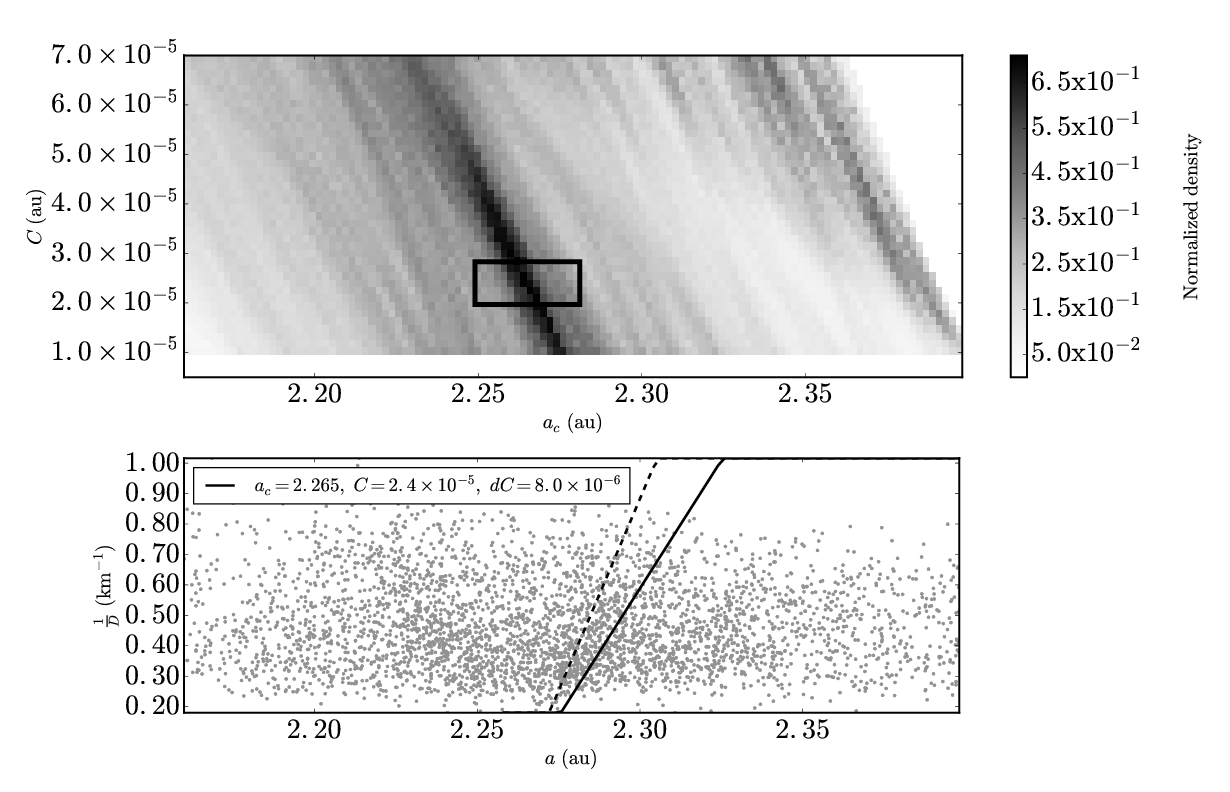}
\else
\includegraphics[scale=0.50]{Baptistina_density.png}
\fi
\else
I am not enabling plots.
\fi
\caption{The density method applied to the Baptistina family V-shape. (Top panel) The normalized density in units of au$^{-1}$ km$^{-1}$ for asteroids in the inner V-shape in the $a_c$-$C$ range, ($a_c\pm \frac{\Delta a_c}{2}$,$C\pm \frac{\Delta C}{2}$) where $\Delta a_c \; = \; 2.0 \times 10^{-3}$ au and $\Delta C \; = \; 1.5 \times 10^{-6}$ au for the Baptistina family. The box marks the peak value in the normalized density for the synthetic family V-shape. (Bottom Panel) $D_r(a,a_c,C,\pv)$ is plotted for the peak values with the primary V-shape as a solid line. The dashed line mark the boundary for the area in $a$ vs. $D_r$ space for the inner V-shape using Eq.~\ref{eqn.apvDvsC},  $D_r(a,a_c,C-dC,\pv)$ where $dC \; = \; 8.0 \x 10^{-6}$ au.}
\label{fig.BaptistinaD}
\end{figure}

\begin{figure}
\centering
\hspace*{-2.7cm}
\ifincludeplots
\ifgrayscale
\includegraphics[scale=0.50]{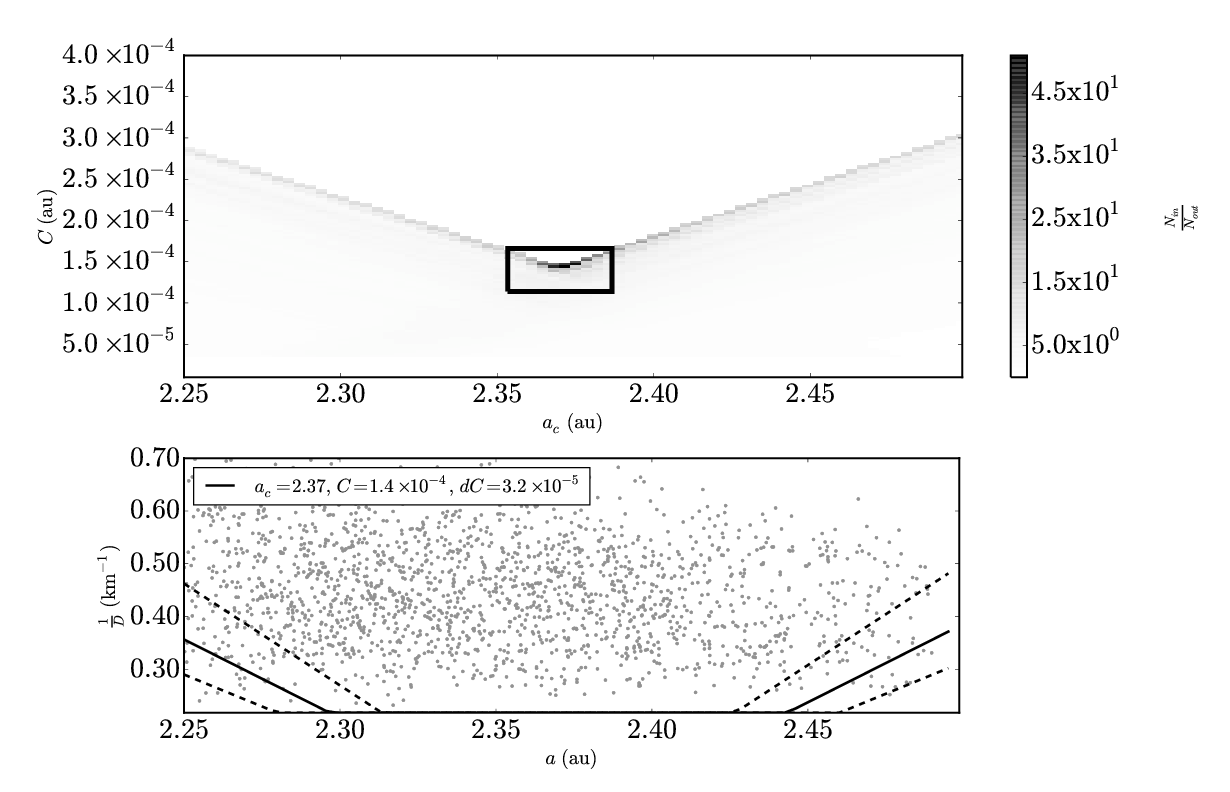}
\else
\includegraphics[scale=0.50]{Vesta_border.png}
\fi
\else
I am not enabling plots.
\fi
\caption{The border method applied to the Vesta family V-shape. (Top panel) The ratio between the number of asteroids in the outer V-shape to the number of asteroids in the inner V-shape in the $a_c$-$C$ range, ($a_c\pm \frac{\Delta a_c}{2}$,$C\pm \frac{\Delta C}{2}$) where $\Delta a_c \; = \; 3.5 \times 10^{-3}$ au and $\Delta C \; = \; 2.7 \times 10^{-6}$ au for Vesta family. The box marks the peak value in $\frac{N_{out}(a_c,C,dC)}{N_{in}(a_c,C,dC)}$ for the synthetic family V-shape. (Bottom Panel) $D_r(a,a_c,C,\pv)$ is plotted for the peak values with the primary V-shape as a solid line where $\pv = 0.05$. The dashed lines mark the boundaries for the area in $a$ vs. $D_r$ space for $N_{in}$ and $N_{out}$ using Eq.~\ref{eqn.apvDvsC},  $D_r(a,a_c,C\pm dC,\pv)$ where $dC \; = \; 3.2 \x 10^{-5}$ au.}
\label{fig.VestaB}
\end{figure}

The same asteroids used to identify the new Polana  family with the V-shape technique were used with the Eulalia family. Eq.~\ref{eq.density_method} are integrated using the interval ($-\infty$,$a_c$] for the Dirac delta function $\delta(a_{j}-a)$ and the interval [$12.0, 17.0$] for the Dirac delta function $\delta( H_j-H )$. A $dC \; = \; 3.2 \; \times10^{-5}$ au was used. Eq.~\ref{eqn.aHvsC} is truncated to 12.0 for $H \; > \; 12.0$ and 17.0 for $H \; > \; 17.0$. The peak in $\rho$ at $(a_c, C) \; = \; (2.49 \; \mathrm{au}, \;8.0 \times 10^{-5} \; \mathrm{au})$ (Fig.~\ref{fig.EulaliaD}, top panel) is similar to the value of $C \; = \; 9.5 \; \times 10^{-5} \; \au$ found in \citet{Walsh2013}. The peak value in $\rho$ found for the Eulalia family is $\sim$5 standard deviations higher than the mean value for $\rho$ in the range 2.0 au $< \; a \;<$ 2.5 au and 5.0 $\times \; 10^{-5}$ au $< \; C \;<$ 4.0 $\times \; 10^{-4}$ au.

The example of using the V-shape technique on the new Polana  and Eulalia families highlights how either the border or density methods are complimentary because each technique is sensitive to finding only one family. Only a peak corresponding to the new Polana  family is detected with the border method (top panel, Fig.~\ref{fig.NewPolanaB}) because the border method is more sensitive to a drop in the number of asteroids in $a$ vs. $D_r$, $H$ space where there are few or no objects outside of the V-shape (bottom panels, Figs.~\ref{fig.NewPolanaB}). Only a peak corresponding to the Eulalia family is found with the density method (top panel, Figs.~\ref{fig.EulaliaD}). The density method is more sensitive to clumps of asteroids, which have a higher density than the background or a family that they are embedded such as in the case of the the Eulalia family being embedded within the new Polana  family (bottom panel, Fig.~\ref{fig.NewPolanaB}).

\subsubsection{Koronis and Karin}

The Koronis and Karin families are examples of families that reside in the same orbital elements space, have similar compositions and albedoes, but have ages that differ by orders of magnitude. The Koronis family is located in the outer main belt between the 5:2 and 7:3 MMRs with Jupiter \citep[][]{Milani1995,Bottke2001}. The Koronis  family consist mostly of S-type members \citep[]{Rivkin2011, Thomas2012} and is $\sim$2 Gyrs old \citep[][]{Broz2013b, Spoto2015}. The Karin family is fully contained within the orbital elements space of the Koronis family and its members have S-type-like $\pv$ of $\sim$0.2 \cite[][]{Harris2009}. The age of the Karin family, 5.8 Myrs, is too young for its members to be dispersed in semi-major axis by the Yarkovsky effect making it an ideal candidate to study family formation events \citep[][]{Nesvorny2002b}. 

\begin{figure}
\centering
\hspace*{-2.7cm}
\ifincludeplots
\ifgrayscale
\includegraphics[scale=0.50]{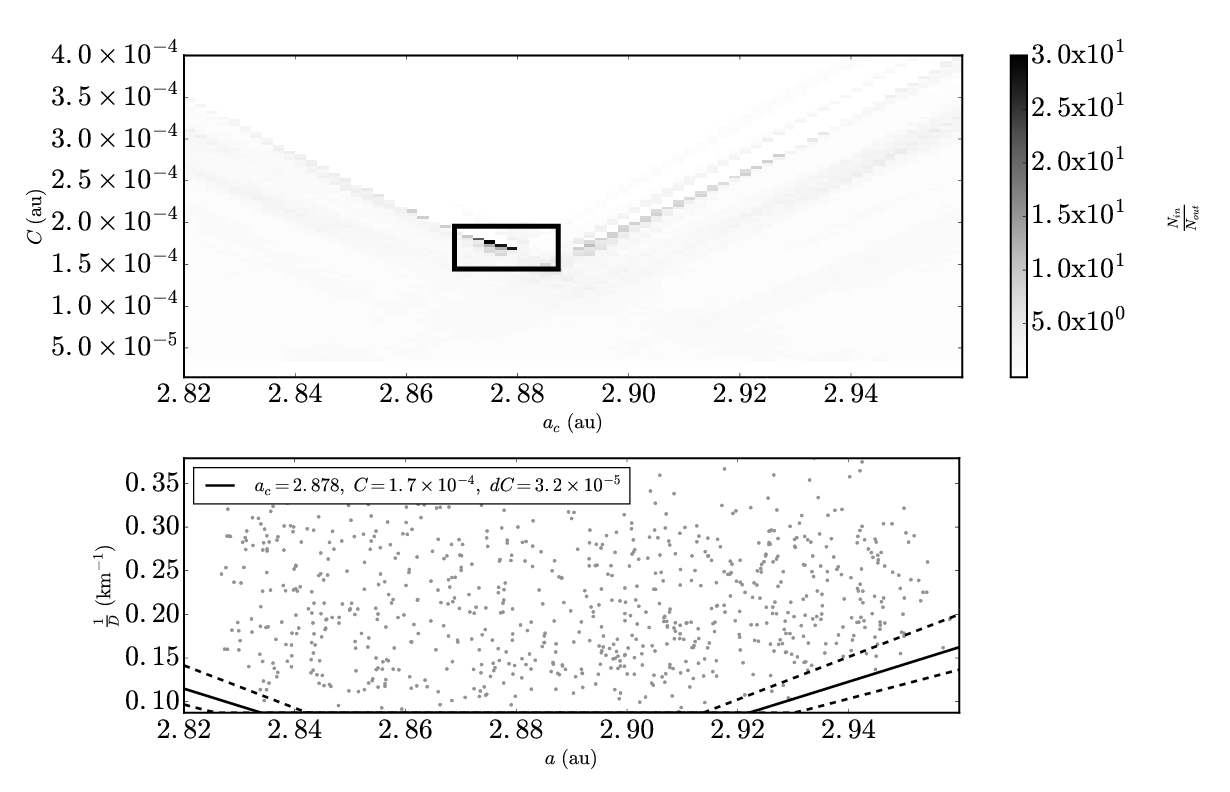}
\else
\includegraphics[scale=0.50]{Koronis_border_grey.png}
\fi
\else
I am not enabling plots.
\fi
\caption{The border method applied to the Koronis family V-shape. (Top panel) The ratio between the number of asteroids in the outer V-shape to the number of asteroids in the inner V-shape in the $a_c$-$C$ range, ($a_c\pm \frac{\Delta a_c}{2}$,$C\pm \frac{\Delta C}{2}$) where $\Delta a_c \; = \; 2.0 \times 10^{-3}$ au and $\Delta C \; = \; 3.7 \times 10^{-6}$ au for Vesta family. The box marks the peak value in $\frac{N_{out}(a_c,C,dC)}{N_{in}(a_c,C,dC)}$ for the synthetic family V-shape. (Bottom Panel) $D_r(a,a_c,C,\pv)$ is plotted for the peak values with the primary V-shape as a solid line where $\pv = 0.2$. The dashed lines mark the boundaries for the area in $a$ vs. $D_r$ space for $N_{in}$ and $N_{out}$ using Eq.~\ref{eqn.apvDvsC},  $D_r(a,a_c,C\pm dC,\pv)$ where $dC \; = \; 3.2 \x 10^{-5}$ au. }
\label{fig.KoronisB}
\end{figure}

The Koronis family V-shape was identified using the border method in $a$ vs. $D_r$ space (Fig.~\ref{fig.KoronisB}) using 765 asteroids with $2.82 \au \; < \; a \; < \; 2.96 \au$, $0.023 \; < \; e \; < \; 0.100$, $0.028 \; < \; \sin i \; < \; 0.045$ \citep[][]{Nesvorny2015a} and $0.2 \; < \; \pv \; < \; 0.5$ \citep[][]{Masiero2013, Spoto2015}. Eqs.~\ref{eq.border_method_N_outer} and \ref{eq.border_method_N_inner} were integrated with the interval ($-\infty$,$\infty$) for the Dirac delta function $\delta(a_{j}-a)$ and the interval [$0.09\;\invkm, 0.38\; \invkm$] was used for the Dirac delta function $\delta(D_{r,j}-D_r )$. Eq.~\ref{eqn.apvDvsC} is truncated to $0.09 \; \invkm$ for $D_r \lesssim 0.09 \; \invkm$ and to $0.38 \; \invkm$ for $D_r \gtrsim 0.38 \; \invkm$. The peak value of  $\frac{N_{in}}{N_{out}}$  found at $a_c \; = \; 2.878$ and $C \; = \; 1.7e-4$, similar to the value of $C \; = \; 2.0 \pm 1.0 \times 10^{-4}$ found by \citet[][]{Nesvorny2015a}. The peak value in the normalized density is $\sim$12 standard deviations above the mean value of  $\frac{N_{in}}{N_{out}}$  in the range 2.82 au $< \; a \;<$ 2.96 au and 1.5 $\times \; 10^{-5}$ au $< \; C \;<$ 4.0 $\times \; 10^{-4}$ au.

\begin{figure}
\centering
\hspace*{-2.7cm}
\ifincludeplots
\ifgrayscale
\includegraphics[scale=0.50]{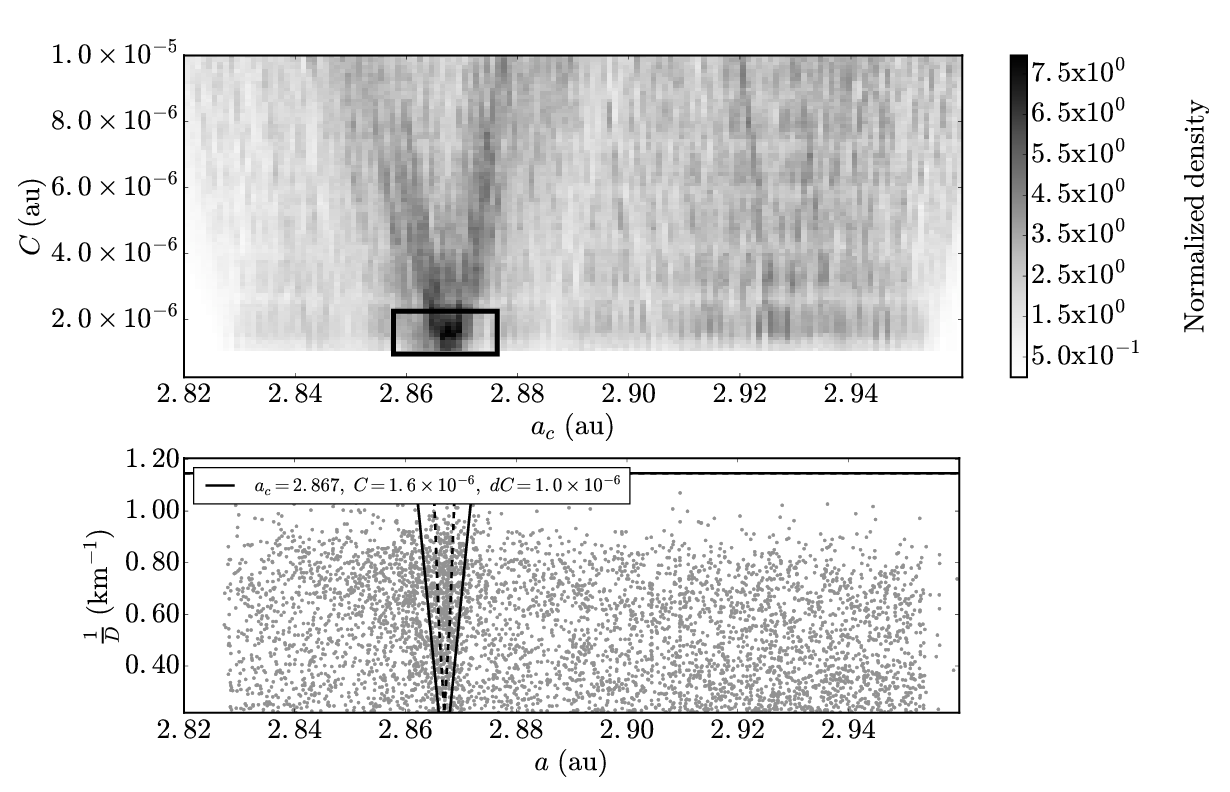}
\else
\includegraphics[scale=0.50]{Karin_density.png}
\fi
\else
I am not enabling plots.
\fi
\caption{The density method applied to the Karin family V-shape. (Top panel) The normalized density in units of au$^{-1} \; \invkm$ for asteroids in the inner V-shape in the $a_c$-$C$ range, ($a_c\pm \frac{\Delta a_c}{2}$,$C\pm \frac{\Delta C}{2}$) where $\Delta a_c \; = \; 1.0 \times 10^{-3}$ au and $\Delta C \; = \; 1.1 \times 10^{-7}$ au for Vesta family. The box marks the peak value in the normalized density for the Karin family V-shape. (Bottom Panel) $D_r(a,a_c,C,\pv)$ is plotted for the peak values with the primary V-shape as a solid line where $\pv \; = \; 0.21$. The dashed line mark the boundary for the area in $a$ vs. $D_r$ space for $\rho$ using Eq.~\ref{eqn.apvDvsC},  $D_r(a,a_c,C- dC,\pv)$ where $dC \; = \; 1.0 \x 10^{-6}$ au.}
\label{fig.KarinD}
\end{figure}

The Karin family was identified with the density method in $a$ vs $D_r$ space (Fig.~\ref{fig.KarinD}) by using 5083 asteroids from the Asteroid Dynamic Site catalogue \citep[][]{Knezevic2003}, in addition to asteroids from the from the \citet[][]{Masiero2011} and \citet[][]{Veres2015} catalogues used in previous real family examples, with proper elements $2.82 \au \; < \; a \; < \; 2.96 \au$, $0.023 \; < \; e \; < \; 0.100$, $0.028 \; < \; \sin i \; < \; 0.045$, the orbital elements ranges that contain the Koronis family \citep[][]{Nesvorny2015a}. Asteroids were limited to $\pv$ range $0.1 \; < \; \pv \; < \; 0.3$, the $\pv$ range of the Karin family \citep[][]{Harris2009}, for asteroids with reliable diameter measurements. Asteroids without diameter measurements were assumed to have a $\pv \; = \; 0.21$, the central $\pv$ value for Karin family members \citep[][]{Harris2009}.

Eq.~\ref{eq.density_method} is integrated using the interval ($-\infty$,$\infty$) for the Dirac delta function $\delta(a_{j}-a)$. The interval [$0.21\;\invkm, 1.20\;\invkm$] was used for the Dirac delta function $\delta(D_{r,j}-H )$. Eq.~\ref{eqn.apvDvsC} is truncated to 0.21$\;\invkm$ for $D_r$ $ <$ 0.21$\;\invkm$ and to 1.20 for $D_r$ $>$ 1.20$\;\invkm$. The peak in $\rho$ is located at $(a_c, C) \; = \; (2.867 \; \mathrm{au}, \;1.6 \times 10^{-6} \; \mathrm{au})$ (Fig.~\ref{fig.KarinD}, top panel) and a $dC \; = \; 1.0 \; \times 10^{-5}$ au was used. The peak value of  $\rho$ found with the density method is $\sim$5 standard deviations above the mean value of  $\rho$  in the range 2.82 au $< \; a \;<$ 2.96 au and 1.0 $\times \; 10^{-6}$ au $< \; C \;<$ 1.0 $\times \; 10^{-5}$ au. The value for $C \; = \; 1.6 \times 10^{-6}$ found with the $density$ for the Karin family V-shape is smaller than the value of $C \; = \; 3\pm1.0 \times 10^{-6}$ from \citet[][]{Nesvorny2015a} possibly due to the density method producing systematically lower values of $C$ compared to values of $C$ constrained with other methods such as the border method as discussed in Section~\ref{S.Ages}.

 The V-shape of the Karin is a direct result of the initial ejection velocities of family fragments due to the parent body's disruption because the Karin family too young to be dispersed in $a$ by Yarkovsky effect \citep[][]{Nesvorny2002b, Harris2009}. The initial ejection velocities of the family fragments is proportional to $\left(\frac{1}{D}\right )^{\beta}$ or a $D_r^\beta$ where $\beta \; = \; 1.0$ for the Karin family \citep[][]{Nesvorny2002b}. The resulting displacement in the fragment's $a$ from their family's V-shape center, $a_c$ caused by the disruption of their parent body is also proportional to $D_r^\beta$. We modify Eq.~\ref{eqn.apvDvsC} to include the variable $\alpha$, for exponent of $D_r$
 
 \begin{equation}
\label{eqn.apvDvsCvsAlp}
D_r(a,a_c,C,\pv,\alpha) \; = \; \left ( \frac{\left | a - a_c \right | \; \sqrt{\pv} }{1329 \km \; C} \right )^{\frac{1}{\alpha}}
\end{equation}
 where $\alpha$ is moved to the right side of the equation. $\alpha \simeq \beta$ in $\delta V \propto \left(\frac{1}{D}\right )^{\beta}$, where $\delta V$ is the initial Velocity of family fragments, for families too young for their fragments to be modified in semi-major axis by the the Yarkovsky effect. The value of $C$ in Eq.~\ref{eqn.apvDvsCvsAlp} describes the width of the V-shape solely due to the spread in fragments caused by the size-dependence of the ejection velocity. 
 
\begin{figure}
\centering
\hspace*{-2.7cm}
\ifincludeplots
\ifgrayscale
\includegraphics[scale=0.50]{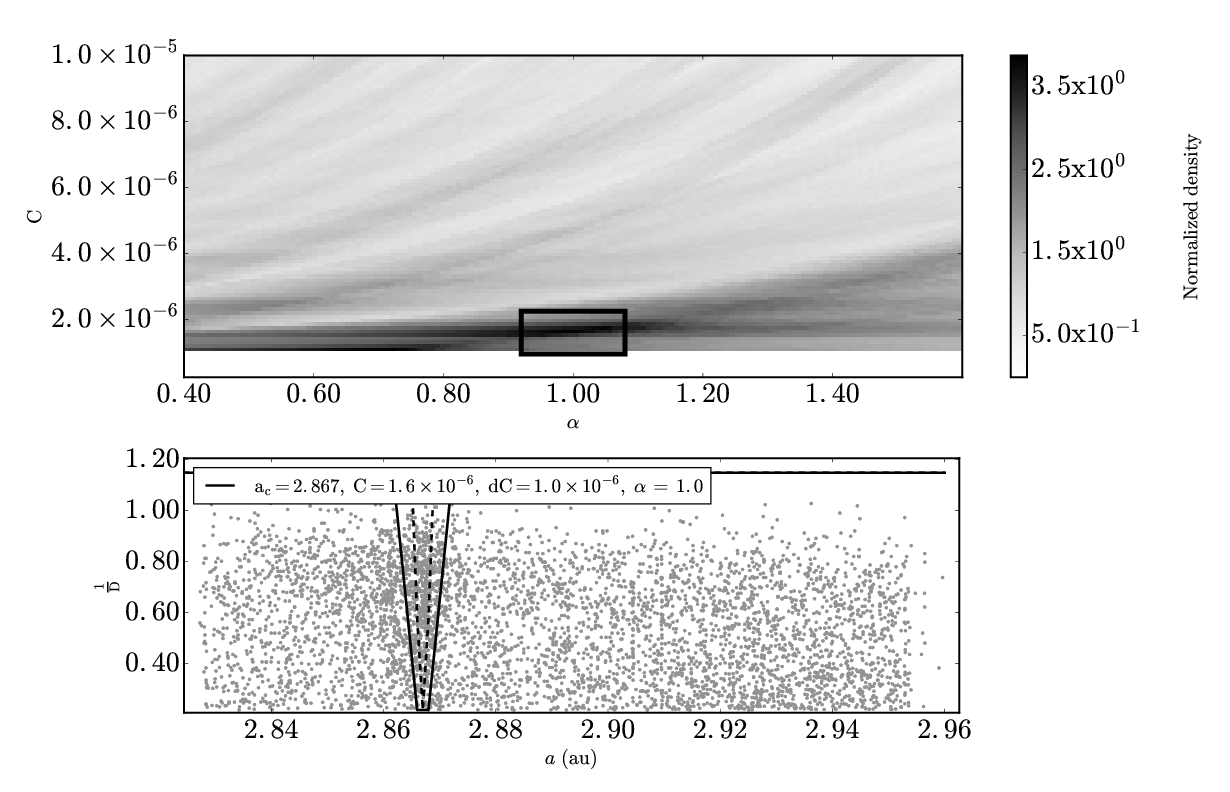}
\else
\includegraphics[scale=0.50]{Karin_alph_v_C_density.png}
\fi
\else
I am not enabling plots.
\fi
\caption{The modified density method applied to the Karin family V-shape. (Top panel) The normalized density in units of au$^{-1}$ km$^{-1}$ for asteroids in the inner V-shape in the $\alpha$-$C$ range, ($\alpha\pm \frac{\Delta \alpha}{2}$,$C\pm \frac{\Delta C}{2}$) where $\Delta \alpha \; = \; 6.1 \times 10^{-3}$ au and $\Delta C \; = \; 1.1 \times 10^{-7}$ au for Vesta family. The box marks the peak value in the normalized density for the Karin family V-shape. (Bottom Panel) $D_r(a,a_c,C, \pv, \alpha)$  is plotted for the peak values with the primary V-shape as a solid line. The dashed line mark the boundary for the area in $a$ vs. $D_r$ space for $\rho$ using Eq.~\ref{eqn.apvDvsCvsAlp},  $D(a,a_c,C- dC,\pv, \alpha)$ where $a_c \; = \; 2.867$ au, $dC \; = \; 1.0 \x 10^{-6}$ au, $\pv \; =  \; 0.21$, the central $\pv$ value for the Karin family \cite[][]{Harris2009}. and $\alpha \; = \; 1.0$.}
\label{fig.KarinD_alph}
\end{figure}

The value of $\alpha$ from Eq.~\ref{eqn.apvDvsCvsAlp}  for the Karin family's V-shape is determined with a modified version of the density method

\begin{equation} 
\label{eq.density_method_alp}
  \rho(a_c,C,dC,\pv,\alpha)
    = \frac{\; \Sigma_j 
       \; w(D_j) \;
       \int\limits_{a_1}^{a_2} da
    	\int\limits_{D_r(a, a_c,C, \pv,\alpha)}^{D_r(a, a_c,C_-, \pv,\alpha)} dD_r
       \; \delta(a_{j}-a)
       \; \delta( D_{r,j}-D_r )}{\int\limits_{a_1}^{a_2}  da
    	\int\limits_{D_r(a, a_c,C,\pv,\alpha)}^{D_r(a, a_c,C_-,\pv,\alpha)} \; dD_r}
\end{equation}
A peak in $\rho$ is found at $\alpha \; = \; 1.0$ and $C \; = \; 1.6 \times 10^{-6}$ (top panel, Fig.~\ref{fig.KarinD_alph}), the same value of $C$ found for the Karin family V-shape when using the unmodified density method. The peak in $\rho$ in the modified density method was found using the same 5083 asteroids from the Asteroid Dynamic Site catalogue with proper elements range and $\pv$ range as used to identify the Karin V-shape with the unmodified method. Eq.~\ref{eq.density_method_alp} is integrated using the interval ($-\infty$,$\infty$) for the Dirac delta function $\delta(a_{j}-a)$. The interval [$0.21\;\invkm, 1.20\;\invkm$] was used for the Dirac delta function $\delta(D_{r,j}-H )$. $a_c$ and $dC$ are fixed to 2.867 au and $dC \; = \; 1.0 \x 10^{-6}$ au respectively. Eq.~\ref{eqn.apvDvsC} is truncated to 0.21$\;\invkm$ for $D_r$ $ <$ 0.21$\;\invkm$ and to 1.20 for $D_r$ $>$ 1.20$\;\invkm$. The peak value of  $\rho$ found with the density method is $\sim$4 standard deviations above the mean value of  $\rho$  in the range 2.82 au $< \; a \;<$ 2.96 au and 0.4 $< \; \alpha \;<$ 1.6. The value of $\alpha \; = \; 1.0$ found with the modified density method matches the results of \citet{Nesvorny2002b}.

\section{Discussion and conclusion}
\label{s.Discussion}
An automated method for identifying collisional asteroid family Yarkovsky V-shapes is demonstrated on synthetic and real collisional families. The V-shape technique is successful at identifying families resulting from catastrophic disruptions and cratering events such as the Erigone and Vesta families respectively. The V-shape technique is successful at detecting families for which there is not total agreement in the literature on their classification as collisional families such as the Flora, which has its family members dispersed on Gyr timescales by resonances and the new Polana  and Eulalia families, which are affected by the close proximity of the 3:1 MMR with Jupiter \citep[][]{Walsh2013}.

Two variations on the V-shape technique were developed, the border and density methods. The border method uses the ratio of the number of objects inside and outside the border of a V-shape and sensitive to collisional families that have a distinct edge with few objects outside their borders. The density method measures the density of objects in $a$ vs. $D_r,H$ space near the edge of a V-shape. The density method underestimates the width of the V-shape by $\sim10\%$ compared to the border method, but is more sensitive to asteroid families embedded in a background of asteroids or other asteroid families.

The V-shape technique was applied with known ranges of proper elements and albedos of known families taken from from the literature \citep[$\eg$,][]{Walsh2013,Dykhuis2014,Masiero2013,Nesvorny2015a}. In addition, a weighting factor, $w(D_j)$, from eqs.~\ref{eq.border_method_N_outer}, \ref{eq.border_method_N_inner} and \ref{eq.density_method} was used assuming all of the known families that the V-shape technique was applied to were created by a catastrophic disruption of their parent body with the exception of the Vesta family. The precise proper element ranges used in the test of the V-shape technique on known families is not as important as long as the ranges include the family's V-shape in $a$ vs. $\frac{1}{D}$,$H$ space. Different weighting factors make the V-shape technique more sensitive to identifying V-shapes of families created by the catastrophic disruption of their parent body, such as the Erigone family \citep[][]{Tanga1999}, versus those created by cratering events such as the Vesta family \citep[][]{Farinella1996}. 

The current V-shape technique can be improved by including asteroid color data from all-sky surveys such as from the Sloan Digital Sky Survey \citep[][]{Ivezic2001} to remove interloping asteroids from V-shapes. Additional MBA diameter measurements such as from the Infrared Astronomical Satellite (IRAS), Midcourse Space Experiment (MSX) and Akari surveys \citep[][]{Tedesco2002a,Tedesco2002b,Usui2011} can be used in addition to the WISE MBA measurements. The IRAS, MSX and Akari surveys include diameter measurements of asteroids that can be used to enhance the V-shape technique because these catalogues include asteroid diameter measurements that are not  in the WISE catalogue. Future surveys such as the Large Synoptic Survey Telescope and Gaia will further enhance the V-shape technique with optical photometry and spectroscopy  \citep[][]{Ivezic2008,Delbo2012, Campins2012,Tanga2016}. The V-shape technique will also benefit from additional optical photometric data of asteorids from ongoing surveys such as PanSTARRS and future optical surveys will enable revised, more accurate absolute magnitude measurements to be made \citep[e.g.,][]{Veres2015}.

As was discussed in Section~\ref{s.Introduction}, the diffusion of proper elements $e$ and $i$ on Gyr timescales may prevent the identification of Gyr-old families by traditional family identification methods. The situation is even more critical for the identification of primordial families, which are families issued from the break-up of asteroids during the early ages of the solar system more than 4 Gyr ago, when the asteroid belt was more populated and the collisional rate was higher. At that time the orbits of the planets were still evolving in a non-periodic way, which should have enhanced the dynamical dispersion of the families. According to current models, the asteroid belt evolved in two stages \citep[see][for a review]{Morbidelli2015a}. The asteroid belt was dynamically excited and severely depleted in the first few million years, possibly due to the existence of resident planetary embryos \citep[][]{Wetherill1992,Petit2001,Obrien2007} or the wide-range migration of Jupiter \citep[][]{Walsh2011}. The identification of asteroid families during this period of time is hopeless due to the fact that the orbital distribution of asteroids was strongly scrambled at that time \citep[][]{Brasil2016}. In the second stage, presumably $\sim 4$~Gyr ago, the orbital distribution in the asteroid belt was shaken again, due a dynamical upheaval of the giant planets \citep[][]{Gomes2005, Morbidelli2010} or a new episode of giant planet migration \citep[][]{Minton2010}. This second phase should have led to the loss of about 50$\%$ of the asteroids still present at the time, as well as to large changes in eccentricities and inclinations; however, only very limited changes should have occurred in semi-major axes, unless a planet temporary invaded the asteroid belt, crossing it for a sufficiently long time \citep[][]{Brasil2016}. In fact, the disturbance of the asteroid belt should have been mostly of ÒsecularÓ nature, related to the change in the eccentricities and inclinations of the major planets and the sweeping of secular resonances. If this vision of the early evolution of the Solar System is correct, the asteroid families formed after the first violent stage, but before or during second stage would be fully dispersed in proper  $e$ and $i$, but would still keep some coherence in semi-major axis \citep[][]{Brasil2016}. Clearly the HCM method and its surrogates would fail in identifying these families. The V-shape method is developed as robust method for finding asteroid families whose fragments have had their proper $e$ and $i$ significantly altered by the stochastic migration of planets during the early age of the solar system.

Future improvements to the V-shape technique will include applying the V-shape finding methods in a search for unknown families covering the entire main belt. The V-shape technique is an ideal tool for finding additional unknown $> \;2$ Gyr-old families because it has been demonstrated as being able to identify families, which are too diffuse or have not been able to be identified with classic methods such as HCM. 

% ACKNOWLEDGEMENTS -----------------------------------------------------------

\acknowledgments

\section*{Acknowledgments}

BTB is supported by l'\`{E}cole Doctorale Sciences Fondatementales et Appliqu\'{e}es, ED.SFA (ED 364) at l'Universit\'{e} de Nice-Sophia Antipolis. KJW was supported by the National Science Foundation, Grant 1518127. BTB and MD also acknowledge support from the French ANR project SHOCKS. The authors also wish to acknowledge the two reviewers of this manuscript, B. Novakovi{\'c} and D. Vokrouhlick{\'y}, for helping to improve this article with their thorough review and valuable suggestions.

% BIBLIOGRAPHY -----------------------------------------------------------

\bibliographystyle{icarus}
\bibliography{VShapeIdentFamilies_v_revised}

% FIGURES -----------------------------------------------------------

\clearpage

\clearpage

\clearpage

\clearpage

\clearpage

\clearpage

\clearpage

\clearpage

\clearpage

\clearpage

\clearpage

\clearpage

\clearpage

\clearpage

\clearpage

\clearpage

\clearpage

\end{document}